\documentclass[aps,prd,twocolumn,nofootinbib,superscriptaddress,tightenlines]{revtex4}
\usepackage[dvipsnames]{xcolor}
\usepackage{amsmath}
\usepackage{dcolumn}
\usepackage{lipsum}
\usepackage{amssymb}
\usepackage{soul}
\usepackage{url}
\usepackage{epsfig}
\usepackage{graphicx}
\usepackage{amsmath}
\usepackage{bm}
\usepackage{setspace}
\usepackage{appendix}
\usepackage{lscape}
\usepackage{amsthm}
\usepackage{bbold}
\usepackage{dcolumn}
\usepackage{epsfig}
\usepackage{graphics}
\usepackage{graphicx}
\usepackage[utf8]{inputenc}

\usepackage{natbib}
\usepackage{graphicx}
\usepackage{dcolumn}
\usepackage{bm}
\usepackage{amsmath}
\usepackage{float}
\usepackage{multirow}
\usepackage{slashed}
\usepackage{xcolor}
\usepackage{physics}
\usepackage{multirow}
\usepackage{gensymb}
\usepackage{mathtools,braket}
\usepackage{subcaption}
\usepackage{lipsum}  
\usepackage{color}
\usepackage{soul}
\usepackage{placeins}
\usepackage[colorlinks=true]{hyperref}
\newcommand{\orcid}[1]{\href{https://orcid.org/#1}{\includegraphics[width=8pt]{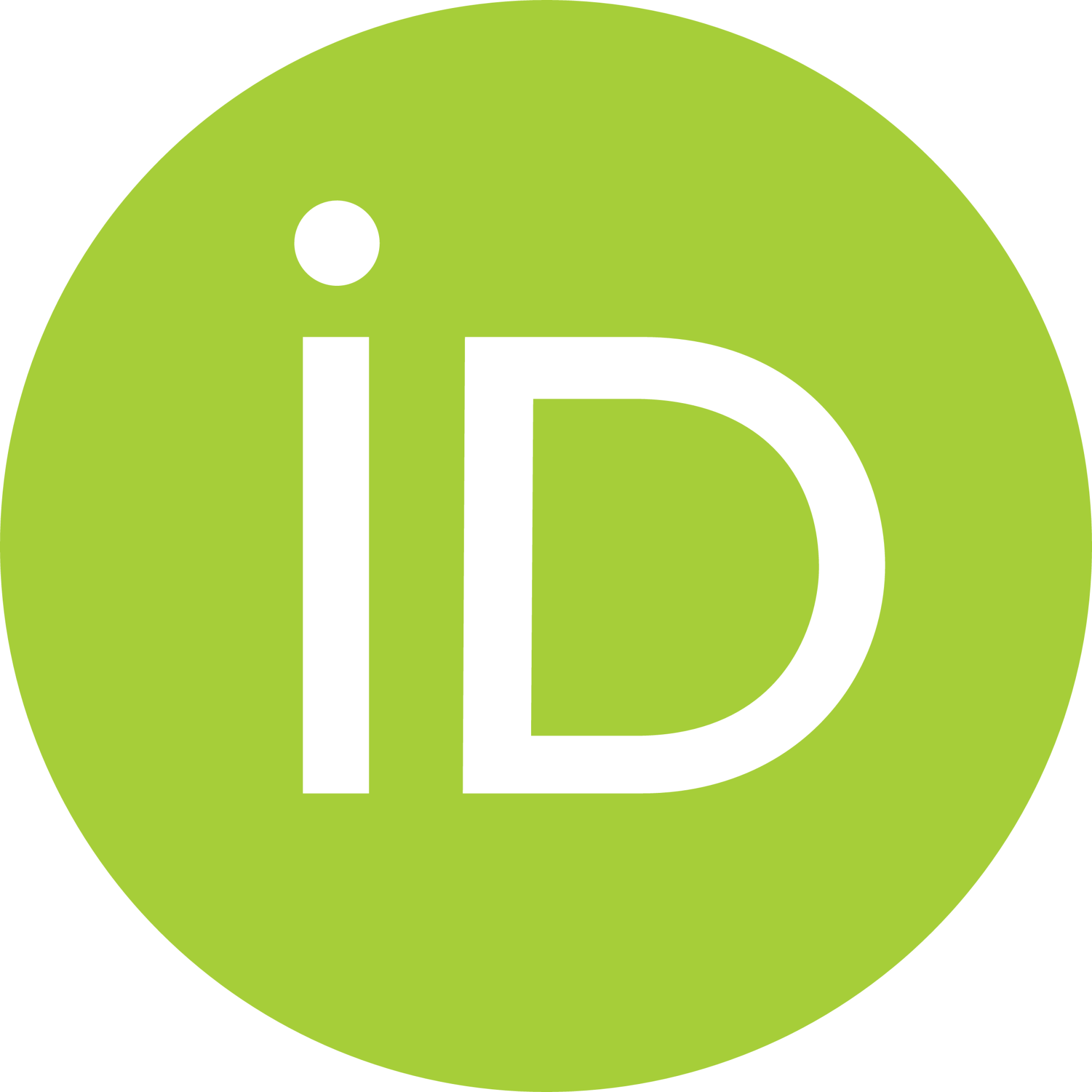}}}

\usepackage{bm}
\usepackage{xspace}
\usepackage{cancel}
\usepackage{float}
\usepackage{multirow}
\definecolor{darkgreen}{rgb}{0,0.5,0}
\definecolor{purple}{rgb}{0.5,0,0.5}
\definecolor{nblue}{rgb}{0.0,0.0,0.50}
\definecolor{scarlet}{rgb}{1.0,0.2,0}
\definecolor{darkmagenta}{rgb}{0.55, 0.0, 0.55}
\definecolor{darkolivegreen}{rgb}{0.33, 0.42, 0.18}
\definecolor{darkcandyapplered}{rgb}{0.64, 0.0, 0.0}

\hypersetup{
    colorlinks=true,
    linkcolor=purple,
    citecolor=purple,
    urlcolor=blue
}
\newcommand{\be}{\begin{equation}}

\newcommand{\ee}{\end{equation}}
\newcommand{\bea}{\begin{eqnarray}}
\newcommand{\eea}{\end{eqnarray}}
\newcommand{\beas}{\begin{eqnarray*}}
\newcommand{\eeas}{\end{eqnarray*}}

\begin{document}
\title{Pion Parton Distribution Functions in the Light-Cone Quark Model and Experimental Constraints}
\author{Hari Govind P\,\orcid{0009-0006-3970-1631}}
\email{harigovinduae@gmail.com}
\affiliation{Computational High Energy Physics Lab, Department of Physics,
Dr.\ B.R.\ Ambedkar National Institute of Technology, Jalandhar, Punjab 144008, India}

\author{Satyajit Puhan\,\orcid{0009-0004-9766-5005}}
\email{puhansatyajit@gmail.com}
\affiliation{Computational High Energy Physics Lab, Department of Physics,
Dr.\ B.R.\ Ambedkar National Institute of Technology, Jalandhar, Punjab 144008, India}
\affiliation{Institute of Physics, Academia Sinica, Taipei 11529, Taiwan}

\author{Abhishek K.~P.\,\orcid{0009-0004-4958-3160}}
\email{abhishekkunjunni@gmail.com}
\affiliation{Computational High Energy Physics Lab, Department of Physics,
Dr.\ B.R.\ Ambedkar National Institute of Technology, Jalandhar, Punjab 144008, India}

\author{Reetanshu Pandey\,\orcid{0009-0007-3524-7766}}
\email{reetanshuhep@gmail.com}
\affiliation{Computational High Energy Physics Lab, Department of Physics,
Dr.\ B.R.\ Ambedkar National Institute of Technology, Jalandhar, Punjab 144008, India}

\author{Harleen Dahiya\,\orcid{0000-0002-3288-2250}}
\email{dahiyah@nitj.ac.in}
\affiliation{Computational High Energy Physics Lab, Department of Physics,
Dr.\ B.R.\ Ambedkar National Institute of Technology, Jalandhar, Punjab 144008, India}

\author{Arvind Kumar\,\orcid{0000-0003-1873-6094}}
\email{kumara@nitj.ac.in}
\affiliation{Computational High Energy Physics Lab, Department of Physics,
Dr.\ B.R.\ Ambedkar National Institute of Technology, Jalandhar, Punjab 144008, India}

\author{Suneel Dutt\,\orcid{0000-0002-0307-2291}}
\email{dutts@nitj.ac.in}
\affiliation{Computational High Energy Physics Lab, Department of Physics,
Dr.\ B.R.\ Ambedkar National Institute of Technology, Jalandhar, Punjab 144008, India}

\begin{abstract}
In this work, we investigate the valence quark parton distribution functions (PDFs) of the pion within the light-cone quark model. The initial quark PDFs are calculated by solving the quark-quark correlation function for the pseudoscalar mesons. The initial quark PDFs have been evolved to higher energy scales through the Dokshitzer–Gribov–Lipatov–Altarelli–Parisi (DGLAP) evolution equations. We also find that our calculated evolved PDFs match experimental and available theoretical extraction data. For the first time, we have also predicted the $F_2$ structure function at next-to-leading (NLO) order accuracy. The calculated $F_2$ structure function has been compared with the available ZEUS and H1 experimental data at DESY-HERA over a wide range of energy scales. Additionally, we display the forward pion production cross-section for the Drell-Yan process caused by pions using the pion PDFs that were calculated and the target nucleon PDFs from the LHAPDF nucleus datasets. The evolved $F_2$ structure function of the pion have been studied at the upcoming electron-ion collider energy kinematics. Overall, it was observed that the quark PDFs of pions computed using the light-cone quark model consistent with the experimental results.

\end{abstract}

\maketitle
\vspace{0.5em}

 \section{Introduction}
\label{intro}
Understanding the complex internal structure of hadrons has always been a challenging task for modern particle and nuclear physicists in quantum chromodynamics (QCD) \cite{Accardi:2012qut,Bacchetta:2006tn,Hughes:1983kf,CTEQ:1993hwr,Gross:2022hyw,Close:1988de,Bloom:1969kc,PHENIX:2004vcz,Marciano:1977su,LHeCStudyGroup:2012zhm,Kovchegov:2012mbw}. Due to the unresolved issues of color confinement and chiral symmetry breaking, direct access to this structure from QCD first principles is still a challenge, along with direct calculations from the fundamental Lagrangian. The hadron structure can be studied through long-distance non-perturbative components from the experimental cross-sections, which are separated from the short-distance perturbative contributions through QCD factorization theorems \cite{Diehl:2003ny,Collins:1989gx,Ji:2004wu,Gardi:2009qi,Ahrens:2010zv,Stewart:2009yx,Izubuchi:2018srq}. The theoretical description of the hadronic structure in the perturbative zone becomes extremely non-trivial due to the complex dynamics of sea quarks, gluons, and valence quarks inside the hadrons. So, the distribution of these quarks and gluons can be studied using the quark-gluon correlation functions in the non-perturbative region through low energy scale models \cite{Mineo:2003vc,Nambu:1961tp,Klevansky:1992qe,Schlumpf:1994bc,Roberts:2000aa,Brodsky:2006uqa,RuizArriola:2002bp}. The different degrees of freedom of the quarks and gluons inside the hadron can be studied using the multi-dimensional distribution functions. These distribution functions are a five-dimensional generalized transverse momentum parton distribution functions (GTMDs) \cite{Meissner:2009ww,Lorce:2011ni,Puhan:2025kzz,Sharma:2024arf}, three-dimensional generalized parton distribution functions (GPDs) \cite{Diehl:2003ny,Belitsky:2005qn,Polyakov:2018zvc,Boffi:2007yc,Guidal:2004nd,Diehl:2004cx,Guidal:2013rya}, three-dimensional transverse momentum parton distribution functions (TMDs) \cite{Boussarie:2023izj,Avakian:2010br,Bacchetta:2024qre,Angeles-Martinez:2015sea,Puhan:2025ujg,Lorce:2011dv}, two-dimensional form factors (FFs) \cite{Puhan:2025pfs,Miller:2010nz,Davoudiasl:2025ifk,Cao:2025dkv} and one-dimensional parton distribution functions (PDFs) \cite{Soper:1996sn,Lai:1994bb,Pumplin:2002vw,Martin:2009iq,Lai:2010vv,Dulat:2015mca,Buckley:2014ana,H1:2009pze,H1:2015ubc,deFlorian:2009vb}. These PDFs constitute one of the most fundamental non-perturbative inputs in QCD, encoding the longitudinal momentum structure of quarks and gluons inside the hadrons. 
\par PDFs describe how a hadron's longitudinal momentum is divided among the quarks and gluons, hence encoding the hadron's non-perturbative structure. One of the primary subjects of hadron physics is the determination of PDFs through the investigation of hard-scattering phenomena. The probability of finding a quark or gluon inside a hadron can be understood in terms of PDFs as functions of the longitudinal momentum fraction $x$. Through the framework of QCD factorization, they offer a crucial link between the underlying partonic dynamics and cross-sections that can be measured experimentally. One can extract the PDFs through long-distance components of the cross-section in deep inelastic scattering (DIS) \cite{Bloom:1969kc,Altarelli:1977zs}, leading neutron electroproduction \cite{H1:2010hym,ZEUS:2002gig} and Drell-Yan processes \cite{Drell:1970wh,COMPASS:2017jbv}. While significant theoretical and experimental studies are happening for the determination of PDFs for the baryons, particularly the nucleons \cite{Lorce:2025aqp,John:2000ur,ATLAS:2021qnl}, comparatively less is known about the partonic structure of mesons. Among the mesonic systems, the pion plays a key role as the lightest quark–antiquark bound state and the pseudo-Goldstone boson associated with the spontaneous breaking of chiral symmetry \cite{Nambu:1961tp}.

\par The pion PDFs have been widely studied using different phenomenological models, such as light-front quantization \cite{Lan:2019rba}, light-front quark model \cite{Choi:2024ptc},  anti-de Sitter (AdS)-QCD model \cite{Kaur:2018ewq,Gutsche:2014zua}, light-front holographic model \cite{deTeramond:2018ecg}, Nambu-Jona-Lasinio (NJL) model \cite{Shigetani:1993dx}, Dyson-Schwinger equations (DSE) model \cite{Shi:2026hqq},  chiral quark model \cite{Broniowski:2007si}, and in Refs. \cite{,Dwibedi:2025vhr,Ghaffarian:2025zeg,Yu:2024ovn,Chen:2024dhz,dePaula:2022pcb,Chang:2021utv,Roberts:2021nhw}. The
pion PDFs have also been investigated within lattice QCD \cite{Miller:2025wgr,Francis:2025pgf,Francis:2025rya,ExtendedTwistedMass:2024kjf}. and in theoretical extractions \cite{Bourrely:2022mjf,Barry:2021osv,Chang:2020rdy,Novikov:2020snp,Sutton:1991ay,Gluck:1999xe,Wijesooriya:2005ir}. However, there is a lack of direct experimental data available for the pion PDFs. The first measurements of pion-induced cross-sections were obtained from the studies of pion structure functions through high-mass muon-pair production at a beam momentum of $225~\mathrm{GeV}/c$ (FNAL-E-0444) at Fermilab~\cite{Newman:1979tv}. In the same period, measurements were also performed at CERN using $\pi^{-}$--Be di-muon production at beam momenta of $150$ and $175~\mathrm{GeV}/c$ (CERN-WA-011)~\cite{Barate:1979da}. Subsequently, the pion structure has been extensively investigated through pion--nucleon Drell--Yan processes in several fixed--target experiments, including CERN-WA-039~\cite{Corden:1980xf}, FNAL-E-0326~\cite{Greenlee:1985gd}, CERN-NA-010~\cite{NA10:1985ibr}, CERN-NA-003~\cite{NA3:1983ejh}, and FNAL-E-0615~\cite{E615:1989bda}. In addition, information on pion PDFs has been extracted from leading--neutron electroproduction measurements at HERA by the ZEUS and H1 collaborations~\cite{ZEUS:2002gig,H1:2010hym}. More recently, pion--induced Drell-Yan data from the COMPASS experiment using a 190-GeV $\pi$ beam have provided a new constraint on the pion structure~\cite{Meyer-Conde:2019frd}. Upcoming electron-ion colliders (EICs) will provide more information about the pion structure functions and PDFs through the Sullivan process \cite{Aguilar:2019teb}.

In this work, we have calculated the valence quark PDFs of the pion by solving the quark-quark correlation functions in the light-cone quark model (LCQM). Being gauge-invariant and relativistic by nature, LCQM is a non-perturbative method. Its primary focus is on valence quarks, which are the essential building blocks that determine the general structure and inherent characteristics of hadrons. For the case of pseudoscalar mesons, there is only a collinear $f(x)$ PDF available at the leading twist, compared to three for the nucleons. The $f(x)$ is the result of the non-flip quark polarizations inside an unpolarized pion. By solving the correlation function, we have derived the PDF in the light-front wave-function (LFWF) form and, further, in the explicit form using the total wave function (spin- and momentum-space wave function). We have solved the quark PDF by using the leading-order meson Fock state, which makes the contributions from the gluon and sea quarks vanish at the initial scale. To compare our PDF with available structure functions and cross-section data, we have performed the evolutions using the Dokshitzer–Gribov–Lipatov–Altarelli–Parisi (DGLAP) equations \cite{Karlberg:2025hxk}. The evolution of the PDFs from low non-perturbative scales to higher momentum scales (through perturbative QCD evolution equations) is necessary to establish a consistent link with phenomenological extractions, lattice QCD results, and the precision measurements of present and future facilities. The valence, gluon, and sea-quark PDFs are calculated at different energy scales and compared with available theoretical extractions. We have also calculated the $F_2$ structure function of the pion at different energy scales at next-to-leading order (NLO) accuracy and matched it to the leading neutron electroproduction data of HERA. We have also calculated the Drell-Yan cross-sections using the pion PDFs obtained from our model, together with the nuclear PDFs taken from the LHAPDF library \cite{Buckley:2014ana}. The resulting predictions are compared with the available experimental data from the E-0615, NA-010, NA-003, WA-070, WA-039, WA-011, and COMPASS experiments, showing overall good agreement within the experimental uncertainties. For the future EIC, we also present predictions for the scale evolution of the pion structure function $F_2$ at different values of $x$.

The paper is organized as follows. In Sec .~\ref {lcqm}, we discuss the LCQM, including the spin and momentum wave functions. In Sec .~\ref {pdf}, we present the results for the pion PDFs, where the explicit forms of the LFWFs and the corresponding quark distributions are derived. Section~\ref{epdf} is devoted to the QCD evolution of the PDFs and the generation of gluon and sea-quark contributions. In Sec.~\ref{sfpdf}, we present the calculations of the pion structure function $F_2 (x,Q^2)$ at NLO accuracy. Section~\ref{crosssection} discusses the computation of pion--induced Drell--Yan cross-sections and their comparison with the available experimental data. Finally, we summarize our findings in Sec .~\ref{conclusion}.
\section{Light-Cone Quark Model}\label{lcqm}
In the LF framework, the hadrons are treated as the bound states of quarks, gluons, and sea-quarks. They are primarily responsible for all the physical and mechanical properties inside the hadrons. The multi-particle Fock-state of a hadron with four vector momenta $P$ can be represented in terms of the momentum and helicity of its constituents as \cite{Lepage:1980fj,Brodsky:1997de,Ji:2003yj,Puhan:2023ekt,Brodsky:2000xy,Brodsky:2000dr,Pasquini:2023aaf,Puhan:2025kzz,Brodsky:1997de}
\begin{equation}
\label{fockstate}
\begin{aligned}
|M(P, S_z) \rangle = & \sum_{n,\lambda_m} \int \prod_{m=1}^n \frac{\mathrm{d} x_m \mathrm{d}^2 \mathbf{k}_{\perp m}}{16\pi^3 \sqrt{x_m}} (16\pi^3) \\
& \times \delta \biggl( 1-\sum_{m=1}^n x_m \biggr) \delta^{(2)} \biggl( \sum_{m=1}^n \mathbf{k}_{\perp m} \biggr) \\
& \times \Psi_{n/m}(x_m, \mathbf{k}_{\perp m}) | n ; x_m P^+, \mathbf{k}_{\perp m}, \lambda_m \rangle .
\end{aligned}
\end{equation}
Here, $|M(P,S_z)\rangle$ denotes the hadron Fock-state with LF four momentum $P=(P^+,P^-,P_\perp)$, and $S_z$ is the spin projection of the hadron. The indices $n$ and $\lambda_m$ denote the number of flavors and the helicities of the m$^{\text{th}}$ constituent, respectively. The helicity $\lambda_m$ will have only the up ($\uparrow$) and down ($\downarrow$) possibilities for quarks. $\Psi_{n/m}(x_m, \mathbf{k}_{\perp m})$ is the LF wave function (LFWFs) of the $^{\text{th}}$ constituent. The four momentum of the m$^{\text{th}}$ constituent is $k_m=(k_m^+,k_m^-,\mathbf{k}_{\perp m})$. In LF dynamics, $k_m^-$ represents the energy, $k_m^+$ represents the longitudinal momentum, and $\mathbf{k}_{\perp m}$ represents the transverse momenta of the m$^{\text{th}}$ constituent. $x_m=k_m^+/P^+$ is the boost-invariant longitudinal momentum fraction carried by the m$^{\text{th}}$ constituent from the parent hadron. Both the longitudinal momentum fractions and the transverse momenta of the constituents satisfy the momentum sum rules
\begin{eqnarray}
   \qquad \sum_{m=1}^n x_m=1, \ \sum_{m=1}^n \mathbf{k}_{\perp m}=0.
\end{eqnarray}
The hadron Fock-state presented in Eq. (\ref{fockstate}) obeys the normalization condition
\begin{equation}
\begin{aligned}
\sum_{n}\sum_{\{\lambda_m\}}
\int
\left[
\prod_{m=1}^{n}
\frac{dx_m\, d^2\mathbf{k}_{\perp m}}{16\pi^3}
\right]
(16\pi^3)\,
\\
\times
\delta\!\left(1-\sum_{m=1}^{n}x_m\right)
\delta^{(2)}\!\left(\sum_{m=1}^{n}\mathbf{k}_{\perp m}\right)
\\
\times
\left|
\Psi_{n/m}(x_m,\mathbf{k}_{\perp m})
\right|^2
=1 .
\end{aligned}
\end{equation}
In this work, we mainly focus on mesons because their Fock-state decompositions are simpler than those of baryons.
The pion, being the lightest meson, can be described as a bound state of quarks, gluons, and sea-quarks as \cite{Pasquini:2023aaf,Ji:2003yj,Kaur:2018ewq,Puhan:2025kzz}
\begin{align}
|M\rangle &= \sum_{q} |q\bar{q}\rangle \, \Psi_{q\bar{q}}
+ \sum_{q,g} |q\bar{q}g\rangle \, \Psi_{q\bar{q}g}
+ \sum_{q,g,g} |q\bar{q}gg\rangle \, \Psi_{q\bar{q}gg} \nonumber \\
&\quad + \sum_{q} |q\bar{q}(q\bar{q})_{\text{sea}}\rangle \,
\Psi_{q\bar{q}(q\bar{q})_{\text{sea}}}
+ \cdots
\label{a1}
\end{align}
As we restrict our analysis of mesons without explicit gluonic and sea-quarks components, the pion Fock-state in Eq.~(\ref{a1}) reduces to $|M\rangle=\sum |q\bar{q}\rangle \, \Psi_{q\bar{q}}$. Neglecting the higher Fock-state contributions, the meson Fock-state is expressed in terms of quark–antiquark helicities at $S_z=0$ (pseudoscalar meson),
\begin{align}
|M(P, S_z=0)\rangle &= \sum_{\lambda_q,\lambda_{\bar q}}\int
\frac{\mathrm{d}x \, \mathrm{d}^2 \mathbf{k}_{\perp}}{\sqrt{x(1-x)}\,2(2 \pi)^3}
\Psi_{q \bar q}(x,\mathbf{k}_{\perp}^2)\, \nonumber\\
&\quad \times
|xP^+,\mathbf{k}_{\perp}, \lambda_q;\, (1-x)P^+,-\mathbf{k}_{\perp},\lambda_{\bar q} \rangle .
\label{meson}
\end{align}
Here, $x$ and $1-x$ are the longitudinal momentum fractions carried by the constituent quark and antiquark and $\lambda_q(\lambda_{\bar q})$ is the quark (antiquark) helicities inside the meson. The four momenta of the constituent quark ($k_q$) and antiquark ($k_{\bar q}$) used in this work are expressed as
\begin{eqnarray}
k_q&\equiv&\bigg(x P^+, \frac{\textbf{k}_\perp^2+m_q^2}{x P^+},\textbf{k}_\perp \bigg),\label{n1}\\
k_{\bar q}&\equiv&\bigg((1-x) P^+, \frac{\textbf{k}_\perp^2+m_{\bar q}^2}{(1-x) P^+},-\textbf{k}_\perp \bigg).
\label{n3}
\end{eqnarray}
The total meson wave function in Eq.$\Psi_{q \bar q}(x,\mathbf{k}_{\perp}^2)$ (\ref{meson}) combines spin and momentum-space components and can be written as \cite{Puhan:2025ujg,Ji:2003yj}
\begin{eqnarray}
\Psi_{q \bar q}(x,\mathbf{k}_{\perp}^2)
  = \mathcal{S}_{S_z}(x,\mathbf{k}_\perp,\lambda_q,\lambda_{\bar q})\,\phi(x,\mathbf{k}^2_\perp).
\label{eq4e}
\end{eqnarray}
Here, $\mathcal{S}_{S_z}(x,\mathbf{k}_\perp,\lambda_q,\lambda_{\bar q})$ represents the spin wave function, while $\phi(x,\mathbf{k}^2_\perp)$ denotes the radial wave function.
For the momentum space wave function in Eq. (\ref{eq4e}), we have considered the Brodsky-Huang-Lepage
(BHL) prescription \cite{Lepage:1980fj, Xiao:2002iv, Xiao:2003wf, Puhan:2023ekt, Puhan:2024jaw, Puhan:2025pfs, Puhan:2025kzz} as 
\begin{align}
\phi(x,\mathbf{k}^2_\perp) =
A \exp \Bigg[
&-\frac{\frac{\mathbf{k}_\perp^2+m_q^2}{x}
+\frac{\mathbf{k}_\perp^2+m_{\bar q}^2}{1-x}}
{8\beta^2}+\frac{m_q^2+m^2_{\bar q}}{4 \beta^2}
\nonumber \\
&-\frac{(m_q^2-m_{\bar q}^2)^2}
{8\beta^2\left(
\frac{\mathbf{k}_\perp^2+m_q^2}{x}
+\frac{\mathbf{k}_\perp^2+m_{\bar q}^2}{1-x}
\right)}
\Bigg] .
\label{bhl-k}
\end{align}
Here, $m_{q (\bar q)}$ are the masses of the quark and antiquark of the meson, respectively. $A$ and $\beta$ are the normalization constant and harmonic scale parameter of the mesons, respectively. 
The normalization constant can be calculated by normalizing the momentum space wave functions, as
\begin{equation}
\int \frac{dx \, d^{2}\mathbf{k}_{\perp}}{2(2\pi)^{3}}
\left| \phi(x,\mathbf{k}^2_{\perp}) \right|^{2}
= 1 .
\end{equation}
$\mathcal{S}_{S_z}(x,\mathbf{k}_\perp,\lambda_q,\lambda_{\bar q})$ in Eq. (\ref{eq4e}) is the front-form spin wave function derived either from the instant form by Melosh-Wigner rotation \cite{Qian:2008px, Xiao:2002iv, Kaur:2020vkq} or by solving the quark-meson vertex with proper Dirac spinors. Both methods yield the same spin-wave functions for the spin-0 pseudo-scalar mesons. So, in this work, we have considered the spin wave function calculated from the quark-meson vertex as done in our previous works \cite{Choi:1996mq,Qian:2008px,Dwibedi:2025vhr,Puhan:2025ujg,Puhan:2025kzz}. The spin wave function can be calculated using the proper vertex for spin-0 pseudoscalar mesons ($S_z=0$) as 
\begin{align}
\mathcal{S}(x,\mathbf{k}_\perp,\lambda_q,\lambda_{\bar q}) =
\bar u(k_q,\lambda_q)
\frac{\mathcal{A}_{q \bar q}\gamma_5}{\sqrt{2(M_{q \bar q}^2-(m_q^2-m_{\bar q}^2))}}
\, v(k_{\bar q},\lambda_2) .
\end{align}
with $\mathcal{A}_{q \bar q}=M_{q \bar q}+m_q+m_{\bar q}$. Here, $u$ and $v$ are the Dirac spinors \cite{Harindranath:1996hq}.  $M_{q \bar q}=\sqrt{\frac{\mathbf{k}^2_\perp+m_q^2}{x}+\frac{\mathbf{k}^2_\perp+m_{\bar q}^2}{1-x}}$ is the bound state mass of the meson. The spin wave function for pseudoscalar mesons ($S_z=0$) with different helicities of quark and antiquark is expressed as \cite{Qian:2008px}
\begin{equation}
\begin{aligned}
\mathcal{S}(x,\mathbf{k}_\perp, \uparrow,\uparrow)
&= \frac{1}{\sqrt{2}}\omega^{-1}(-\mathbf{k}^{L})
\mathcal{A}_{q \bar q}, \\
\mathcal{S}(x,\mathbf{k}_\perp, \uparrow,\downarrow)
&= \frac{1}{\sqrt{2}}\omega^{-1}\big((1-x)m_q+x m_{\bar q}\big)
\mathcal{A}_{q \bar q}, \\
\mathcal{S}(x,\mathbf{k}_\perp, \downarrow,\uparrow)
&= \frac{1}{\sqrt{2}}\omega^{-1}\big(-(1-x)m_q-x m_{\bar q}\big)
\mathcal{A}_{q \bar q}, \\
\mathcal{S}(x,\mathbf{k}_\perp, \downarrow,\downarrow)
&= \frac{1}{\sqrt{2}}\omega^{-1}(-\mathbf{k}^{R})
\mathcal{A}_{q \bar q} .
\label{spin1}
\end{aligned}
\end{equation}
with $\omega=\mathcal{A}_{q \bar q}\sqrt{x(1-x)[M_{q \bar q}^2-(m_q-m_{\bar q})^2]}$. The above spin wave function obeys the normalization conditions 
\begin{equation}
\sum_{\lambda_q,\lambda_{\bar q}}
\left|
\mathcal{S}(x,\mathbf{k}_\perp,\lambda_q,\lambda_{\bar q})
\right|^{2}
=1 .
\end{equation}
Now, finally, the two-particle Fock-state for pseudoscalar mesons in Eq. (\ref{meson}) can be written with all possible helicities of its constituent quark and antiquark, along with the momentum space wave function, which can be written as
\begin{widetext}
\begin{equation}
\begin{aligned}
|M(P,S_z=0)\rangle
=&
\int
\frac{dx\, d^2\mathbf{k}_\perp}
{2(2\pi)^3 \sqrt{x(1-x)}}
\,\phi(x,\mathbf{k}_\perp^{2})
\\
&\times
\Big[
\mathcal{S}(x,\mathbf{k}_\perp,\uparrow,\uparrow)
|xP^+,\mathbf{k}_\perp,\uparrow;\,(1-x)P^+,-\mathbf{k}_\perp,\uparrow\rangle+
\mathcal{S}(x,\mathbf{k}_\perp,\downarrow,\downarrow)
|xP^+,\mathbf{k}_\perp,\downarrow;\,(1-x)P^+,-\mathbf{k}_\perp,\downarrow\rangle
\\
&\quad+
\mathcal{S}(x,\mathbf{k}_\perp,\downarrow,\uparrow)
|xP^+,\mathbf{k}_\perp,\downarrow;\,(1-x)P^+,-\mathbf{k}_\perp,\uparrow\rangle+
\mathcal{S}(x,\mathbf{k}_\perp,\uparrow,\downarrow)
|xP^+,\mathbf{k}_\perp,\uparrow;\,(1-x)P^+,-\mathbf{k}_\perp,\downarrow\rangle
\Big] .
\label{eqeq}
\end{aligned}
\end{equation}
\end{widetext}

\section{Parton Distribution Functions}
\label{pdf}
The probability of finding the valence quark in a pion with a longitudinal momentum fraction $x$ can be accessed through the PDFs. For the case of the pseudo-scalar mesons, there is only a single unpolarized quark PDF present at the leading twist compared to three for spin-$\frac{1}{2}$ nucleons and four for spin-1 mesons. At a fixed light-front time $\tau$, the quark-quark correlator of the PDF is defined as \cite{Maji:2016yqo}
\begin{equation}
\begin{aligned}
f(x)
&=
\frac{1}{2}
\int \frac{dz^-}{4\pi}\,
e^{\, i k^{+} z^{-}/2}
\\
&\times
\Big\langle M(P,S_z=0)\Big|
\bar{\psi}(0)\, w(0,z)\,\Gamma\,\psi(z)
\Big|M(P,S_z=0)\Big\rangle
\\
&\Big|_{\, z^{+}=0,\;\mathbf{z}_{\perp}=0} .
\end{aligned}
\end{equation}
Here, $\Gamma=\gamma^+$ is the LF vector current for the unpolarized quark PDFs inside the mesons, which also determines the Lorentz structure of the correlator. $\psi(z)$ is the quark field operator. $z=(z^+,z^-,z_\perp)$ is the position four vector,  which is the path of the quark field operators. The Wilson line $w(0,z)$ preserves the gauge invariance of the bilocal quark field operators in the correlation functions \cite{Bacchetta:2020vty} and determines the path of the quark field operators, which has been taken as unity here.
\par Now using the meson Fock-state of Eq. (\ref{eqeq}) and quark field operators, the overlap form of the unpolarized PDF $f(x)$ is found to be,
\\
\\
\\
\\
\\
\begin{equation}
\begin{aligned}
f(x)
&=
\int \frac{d^{2}\mathbf{k}_\perp}{16\pi^{3}}
\, |\phi(x,\mathbf{k}_\perp^{2})|^{2}
\\
&\times
\Big[
|\mathcal{S}(x,\mathbf{k}_\perp,\uparrow,\uparrow)|^{2}
+
|\mathcal{S}(x,\mathbf{k}_\perp,\downarrow,\downarrow)|^{2}
\\
&+
|\mathcal{S}(x,\mathbf{k}_\perp,\downarrow,\uparrow)|^{2}
+
|\mathcal{S}(x,\mathbf{k}_\perp,\uparrow,\downarrow)|^{2}
\Big] .
\end{aligned}
\end{equation}
Now, using the spin wave functions Eq. (\ref{spin1}), the explicit form of the quark PDF is found to be 

\begin{equation}
f(x) =\int \frac{d^{2}\mathbf{k}_\perp}{16\pi^{3}}
\, |\phi(x,\mathbf{k}_\perp^{2})\, \mathcal{A}_{q \bar q}|^2\frac{\textbf{k}_\perp^2+\Big((1-x)m_q+x m_{\bar q}\Big)^2}{\omega^2}.
\end{equation}
\begin{figure}[t]
    \centering
    \includegraphics[width=\columnwidth]{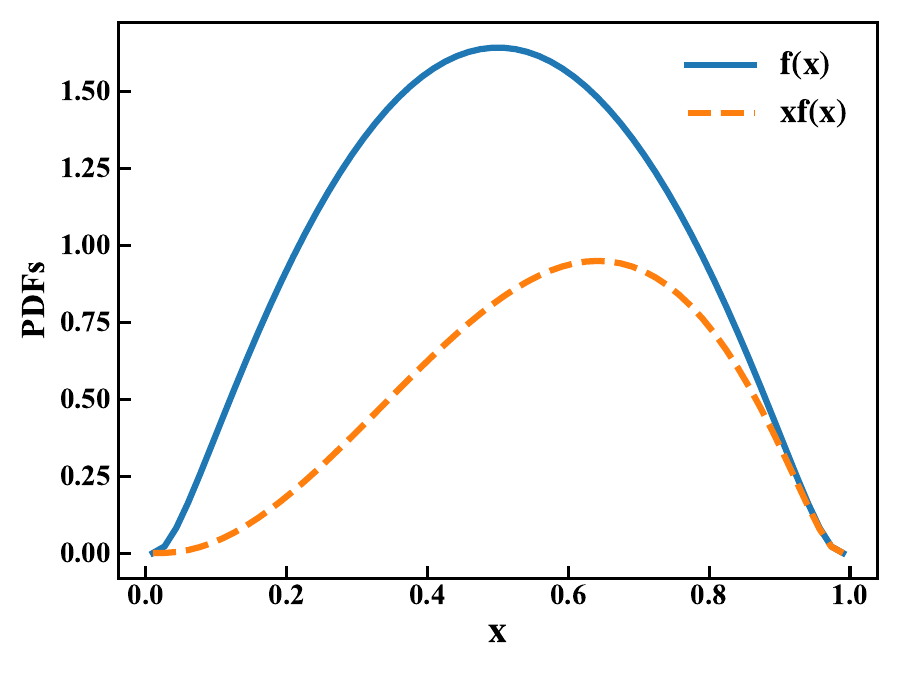}
    \caption{
    The unpolarized pion parton distribution function $f(x)$ 
    and the momentum-weighted distribution $x f(x)$ 
    obtained within the LCQM at the model scale. 
    }
    \label{fig:pion_pdf}
\end{figure}
\begin{figure}[ht]
    \centering
    \includegraphics[width=\columnwidth]{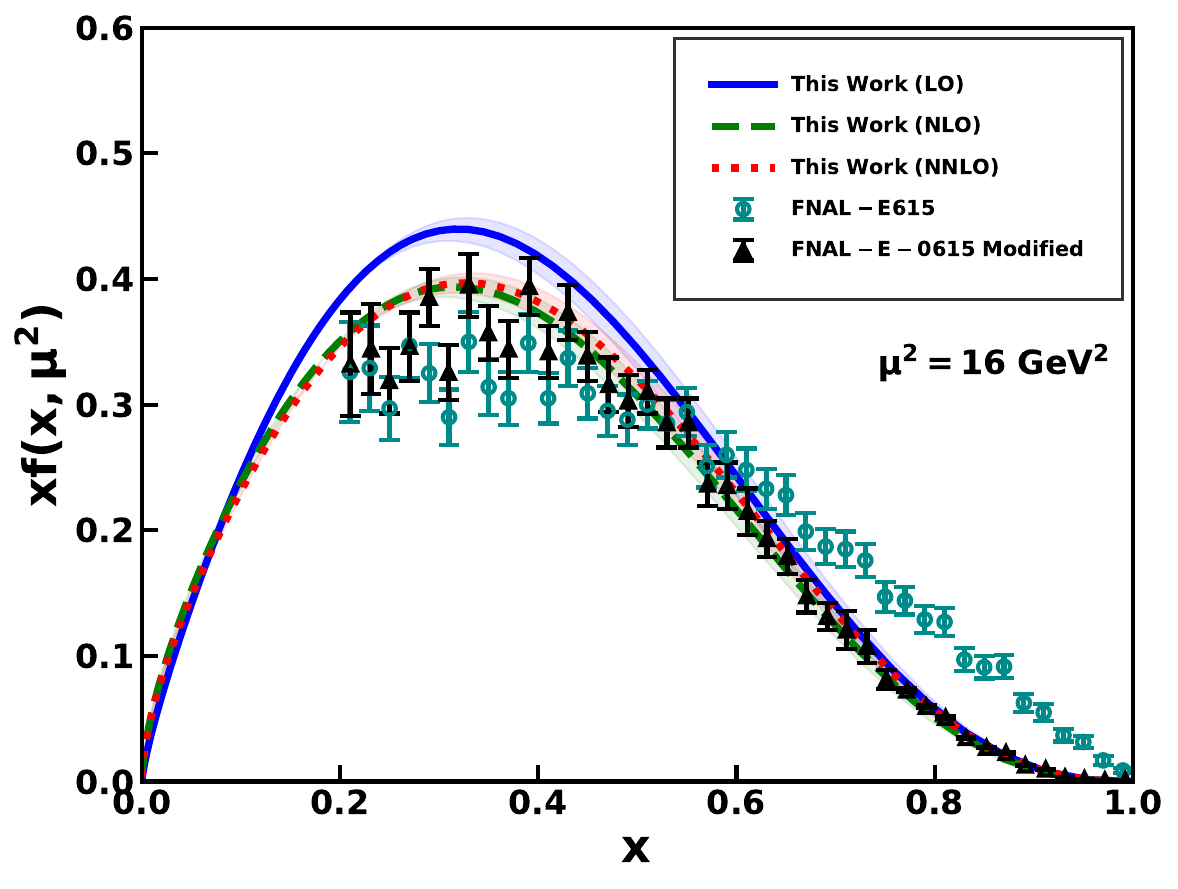}
    \caption{The quark PDF $xf(x)$ as a function of $x$ for the pion compared with FNAL-E-0615 \cite{E615:1989bda} and modified FNAL-E-0615 \cite{Aicher:2010cb} experimental results.}
    \label{fig:pion_pdf_evolution}
\end{figure}
\begin{figure*}[t]
    \centering
    \begin{subfigure}{0.32\textwidth}
        \centering
        \includegraphics[width=\linewidth]{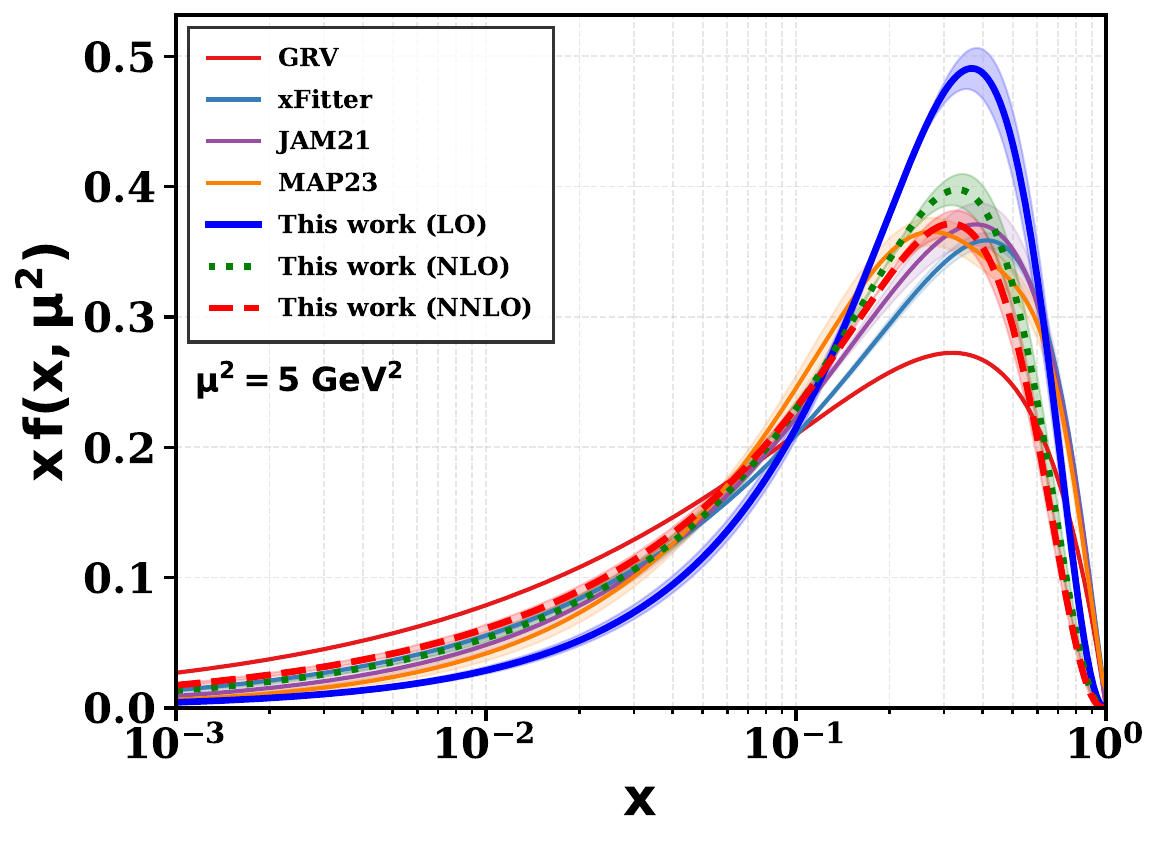}
        \caption{}
        \label{fig:valence}
    \end{subfigure}
    \hfill
    \begin{subfigure}{0.32\textwidth}
        \centering
        \includegraphics[width=\linewidth]{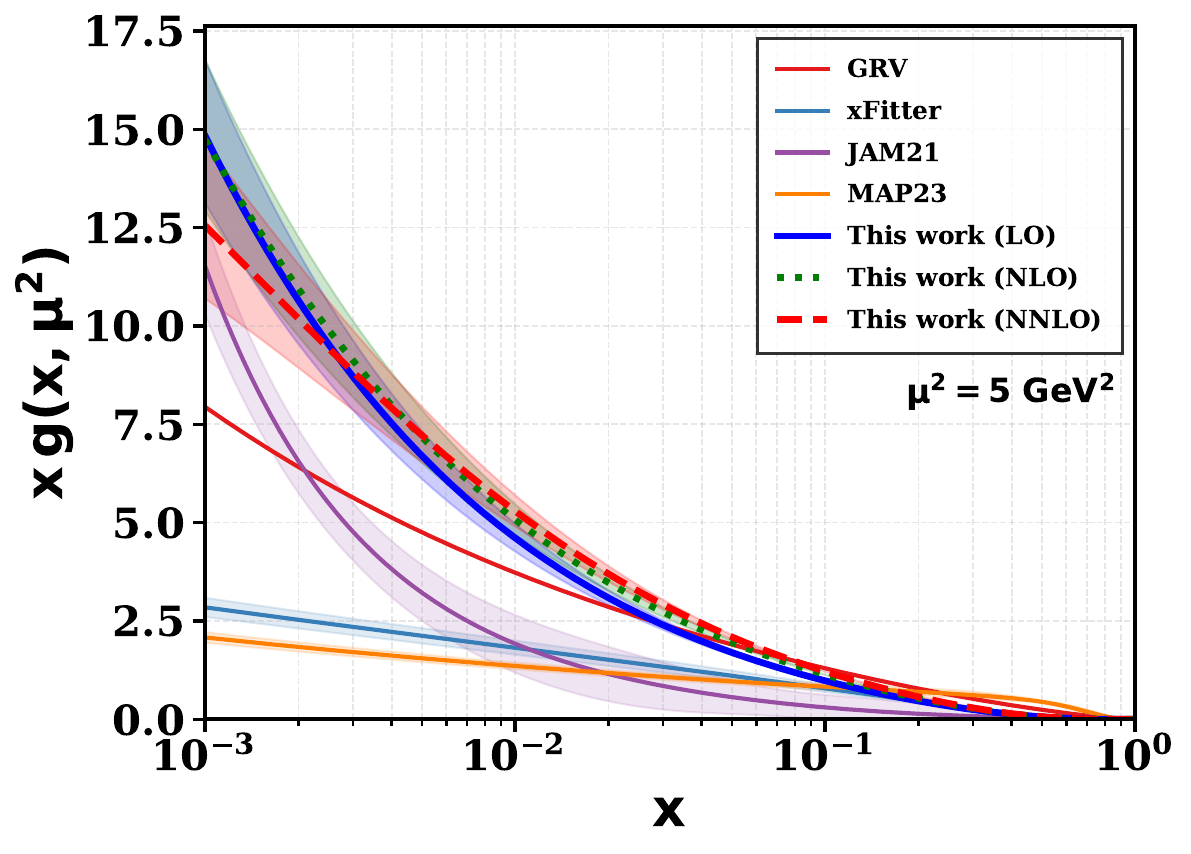}
        \caption{}
        \label{fig:gluon}
    \end{subfigure}
    \hfill
    \begin{subfigure}{0.32\textwidth}
        \centering
        \includegraphics[width=\linewidth]{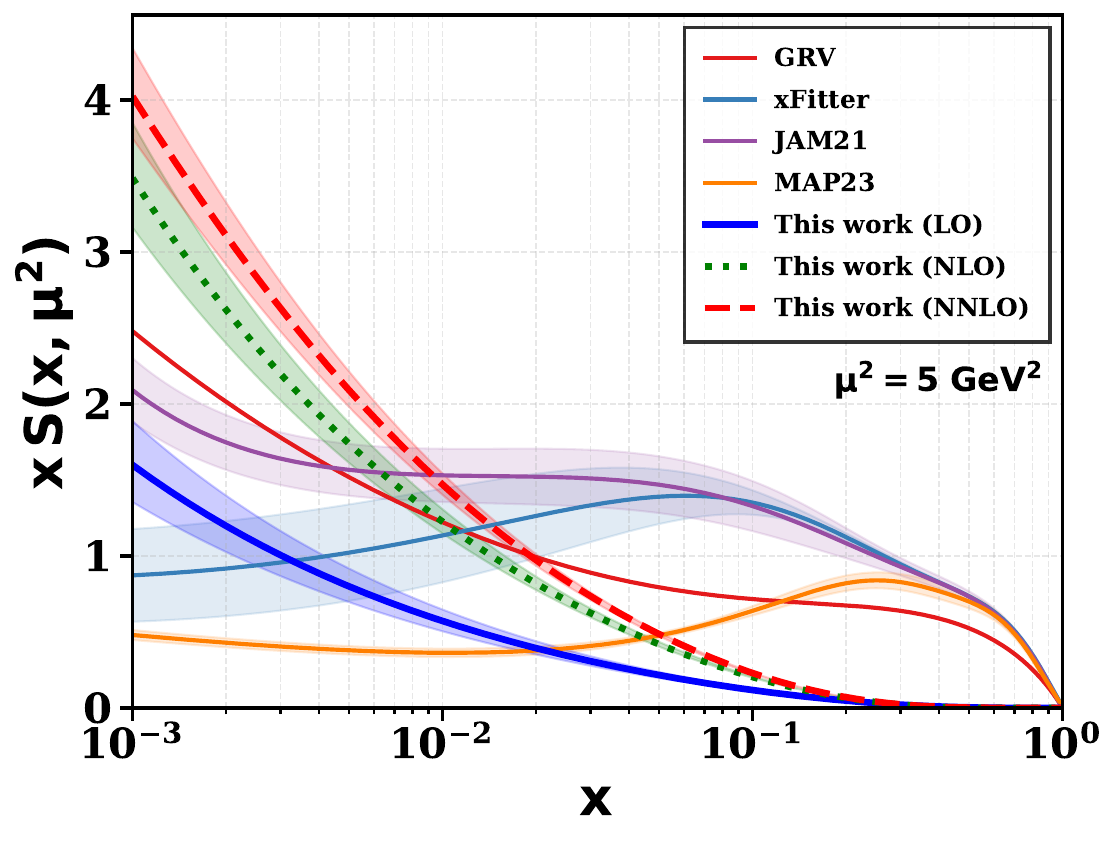}
        \caption{}
        \label{fig:sea}
    \end{subfigure}

    \caption{The evolved (a) valence quark, (b) gluon, and (c) sea-quark PDFs obtained at LO, NLO, and NNLO are compared with the available theoretical extractions from GRV \cite{Gluck:1991ey}, xFitter \cite{Novikov:2020snp}, JAM \cite{Barry:2021osv}, and MAP \cite{Pasquini:2023aaf} at the scale $\mu^{2}=5~\mathrm{GeV}^{2}$, using the initial scale $\mu_{0}=0.6\pm 0.1~\mathrm{GeV}$.}
    \label{fig:pion_comparison}
\end{figure*}
\begin{figure*}[t]
\centering

\begin{subfigure}[t]{0.48\textwidth}
\centering
\includegraphics[width=\linewidth]{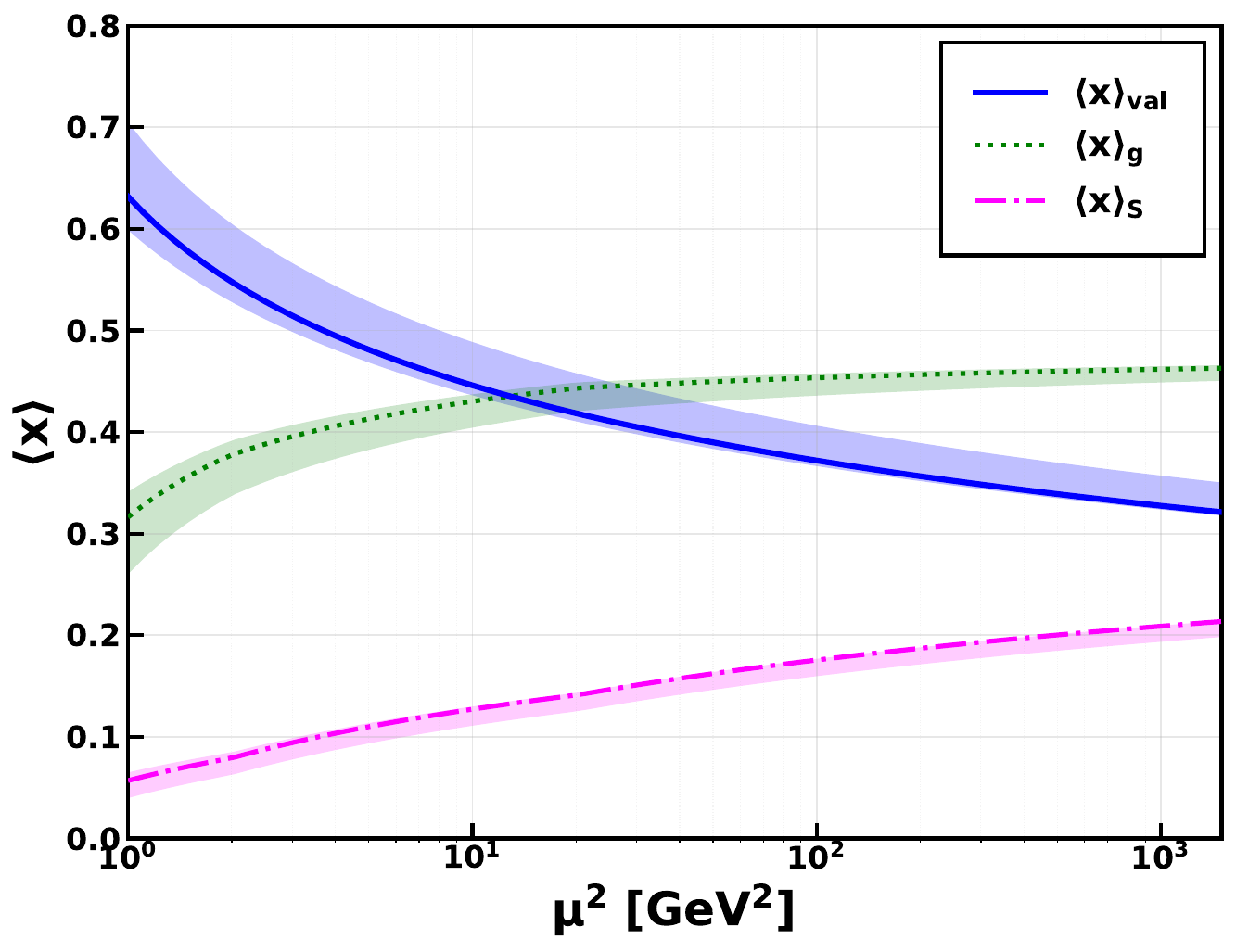}
\caption{}
\end{subfigure}
\hfill
\begin{subfigure}[t]{0.48\textwidth}
\centering
\includegraphics[width=\linewidth]{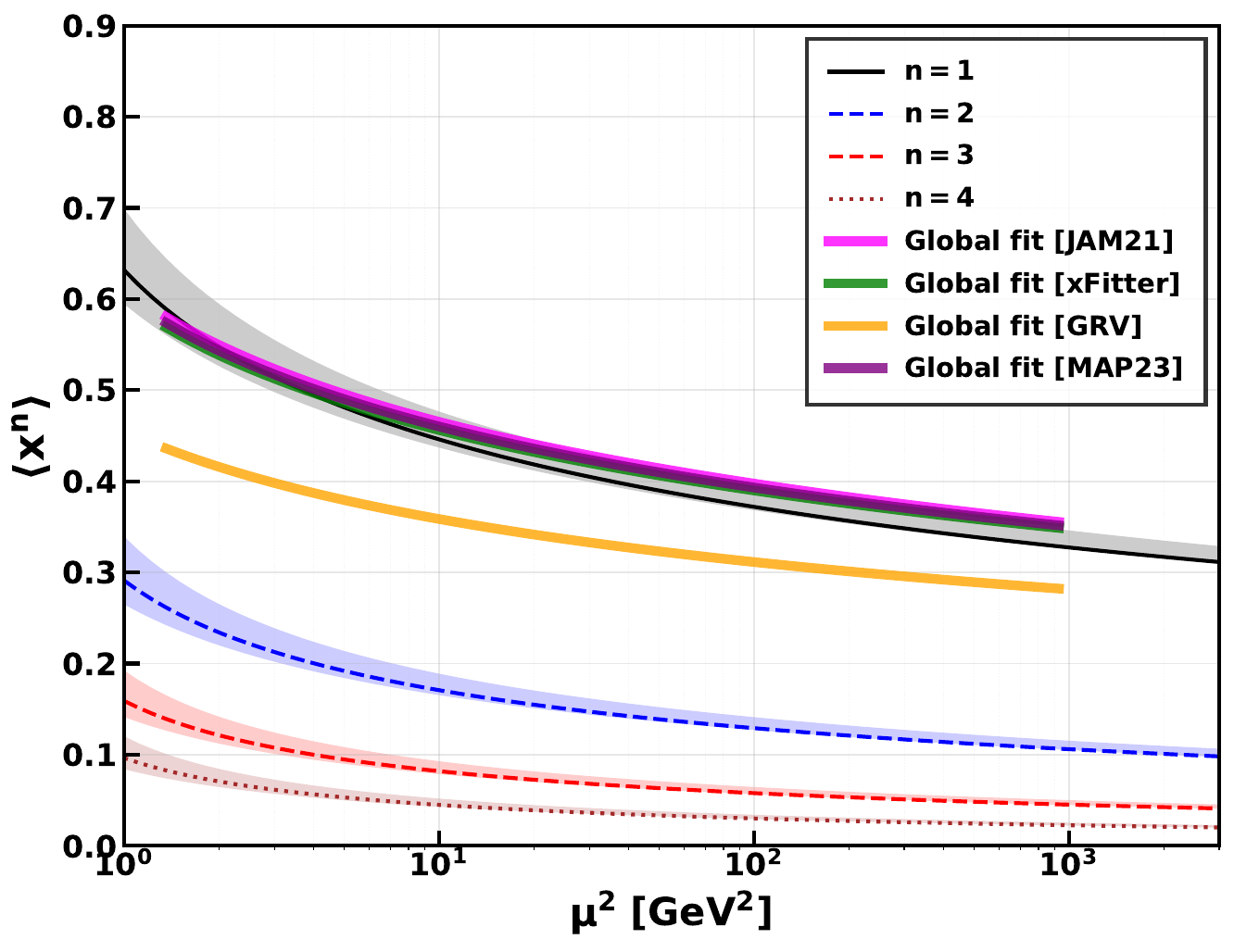}
\caption{}
\end{subfigure}

\caption{(Color online) (a) The average Mellin moments carried by the valence quarks, gluons, and sea-quarks as a function of $\mu^2$. (b) The $n^{th}$ Mellin moment of the valence quark and antiquark $\langle x^n \rangle$ as a function of $\mu^2$, along with comparison with GRV \cite{Gluck:1999xe}, xFitter \cite{Novikov:2020snp}, JAM \cite{Barry:2021osv}, and MAP \cite{Pasquini:2008ax} theoretical extractions data.}
\label{fig:kaon_results}
\end{figure*}
One can obtain the antiquark PDF $\bar f(x)$ by using the momentum sum rule as $f(1-x)$. However, due to the equal mass of $u$ and $d$ antiquarks of the pion, both quark and antiquark PDFs are equal. The unpolarized quark PDF obeys the PDF sum rule \cite{Kaur:2020vkq,Puhan:2023ekt,Puhan:2023hio}
\begin{equation}
\begin{aligned}
\int_{0}^{1} dx\, f(x,\mu_0^2) &= 1 , \\
\int_{0}^{1} dx\, \bar f(x,\mu_0^2) &= 1 , \\
\int_{0}^{1} dx\, x\,\big[f(x,\mu_0^2)+\bar f(x,\mu_0^2)\big] &= 1 , \\
\int_{0}^{1} dx\, g(x,\mu_0^2) &= 0 , \\
\int_{0}^{1} dx\, S(x,\mu_0^2) &= 0 . \\
\end{aligned}
\end{equation}
Here, $g$ and $S$ denote gluons and sea quarks, respectively. However, in this work, we have not considered gluon and sea-quark contributions at the initial scale, therefore, the total momentum of the pion will be equally distributed between the quark and antiquark. The unpolarized quark PDF is found to be the result of the non-flip quark polarizations inside the pion.
\par For the numerical calculations, we have considered equal masses for quark antiquark, i.e., $m_q = m_{\bar{q}} =$ 0.20 GeV. Also, the harmonic scale parameter $\beta$ for pion is taken as 0.410 GeV. These parameters have been adopted from our previous works \cite{Puhan:2023ekt,Puhan:2023hio,Puhan:2025kzz}, where these are calculated by fitting with the mass of the pion. The decay constant of the pion is found to be $122.9$ MeV, close to the particle data group (PDG) value of $130.2$ MeV \cite{ParticleDataGroup:2022pth}. In Fig. \ref{fig:pion_pdf}, we have plotted the quark PDF $f(x)$ and $xf(x)$ as a function of longitudinal momentum fraction $x$ carried by the active quark from the pion at the initial scale. The quark PDF ($f(x)$) is found to be symmetric under the transformation $x \leftrightarrow (1-x)$ due to the equal masses for quark and antiquark, while the $xf(x)$ is found to have maximum distribution around $x > 0.5$. The antiquark $x \bar f(x)$ is found to have opposite distributions to that of $xf(x)$. We have also calculated the lower and higher Mellin moments $\langle x^n \rangle$ of the PDF, which can be computed from
\begin{eqnarray}
    \langle x^n \rangle= \frac{\int dx x^{n} f(x,\mu^2)}{\int dx  f(x,\mu^2)}.
\end{eqnarray}
\begin{table}[h]
\centering
\begin{tabular}{|c|c|c|c|c|c|c|}
\hline
$n$ & 0 & 1 & 2 & 3 & 4 & 5 \\
\hline
$\langle x^n \rangle$ & 1.00 & 0.50 & 0.29 & 0.19 & 0.13 & 0.09 \\
\hline
\end{tabular}
\caption{Mellin moments of the pion PDF at the initial scale.}
\label{mellin}
\end{table}
The $n^{th}$ Mellin moments of the PDF at the initial scale $\mu_0$ have been presented in Table \ref{mellin}. At $n=0$, the Mellin moment provides the information about the number of valence quarks, which is found to be 1. While for $n=1$, which tells us about the average momentum fraction carried by the constituent, is found to be $0.5$ indicating equal momentum distribution among quark and antiquark inside the pion. The higher Mellin moments of the pion are found to be consistent with Ref. \cite{Zhang:2023oja}. Studying PDFs within non-perturbative models provides essential insight into the intrinsic structure of hadrons at low energy scales. But for significant phenomenological applications, this is insufficient on its own. These model-generated PDFs must be evolved to higher momentum scales using perturbative QCD evolution equations to provide solid predictions for upcoming high-energy experiments and to enable trustworthy comparisons with current experimental measurements. This scale evolution connects the perturbative regime studied in deep inelastic scattering and collider processes to the non-perturbative dynamics governing hadron structure at low scales.

\section{Evolution of Parton Distribution Functions}
\label{epdf}
We perform the QCD evolution of our initial scale valence PDF to the relevant experimental scales $\mu^2 = 5$ and $   16\text{ GeV}^2$ with independently adjustable initial scales of the pion using DGLAP equations. We use the higher-order perturbative parton evolution toolkit HOPPET \cite{Karlberg:2025hxk} to solve the DGLAP equations numerically. The first step is to calculate the initial scale of our PDFs by fitting with the available experimental results. To find the initial scale, we evolve our PDFs to $16$ GeV$^2$ and fit with modified FNAL-E-0615 data in Fig. \ref{fig:pion_pdf_evolution} through next-to-next-to-leading order (NNLO) DGLAP evolutions. We have calculated the initial scale as $\mu_0=0.6 \pm 0.1$ with the $\chi^2$ per degree of freedom (d.o.f.) of $1.51$. Throughout this work, we employ the above initial scale for all calculations. 
\begin{table}[t]
\centering
\caption{Initial scales and the $\chi^2/(\mathrm{d.o.f.})$ at LO, NLO, and NNLO obtained from the DGLAP evolution. The $\chi^2$ values are calculated with respect to the central values of the modified FNAL-E-0615 data \cite{Aicher:2010cb}.}
\label{tab:initial_scales}
\begin{tabular}{ccc}
\hline\hline
Order & $\mu_0$ (GeV) & $\chi^2/(\mathrm{d.o.f.})$ \\
\hline
LO   & $0.6 \pm 0.1$ & 4.82 \\
NLO  & $0.6 \pm 0.1$ & 1.89 \\
NNLO & $0.6 \pm 0.1$ & 1.51 \\
\hline\hline
\end{tabular}
\end{table}
In Fig.~\ref{fig:pion_pdf_evolution}, we present the LO, NLO, and NNLO evolved PDFs, and compare them with the FNAL-E-0615 data~\cite{E615:1989bda} as well as the modified FNAL-E-0615 analysis~\cite{Aicher:2010cb}. The corresponding values of $\chi^{2}/(\mathrm{d.o.f.})$ for each perturbative order, obtained using the modified FNAL-E-0615 data, are listed in Table~\ref{tab:initial_scales}. The LO-evolved PDF exhibits a comparatively higher peak than the NLO and NNLO results. Furthermore, our evolved PDFs show noticeable deviations from the FNAL-E-0615 data in the large-$x$ region. This observation agrees with perturbative QCD calculations \cite{Yuan:2003fs}.
\par The DGLAP equation, which bridges PDFs between a final scale and an initial scale is given by,
\begin{align}
\frac{\partial f(x,\mu^{2})}{\partial \ln \mu^{2}}
&=
\frac{\alpha_s(\mu^{2})}{2\pi}
\left[
P_{qq}\otimes q + P_{qg}\otimes g
\right],
\\
\frac{\partial g(x,\mu^{2})}{\partial \ln \mu^{2}}
&=
\frac{\alpha_s(\mu^{2})}{2\pi}
\left[
P_{gq}\otimes q + P_{gg}\otimes g
\right],
\end{align}
with
\begin{equation}
(P \otimes f)(x)=\int_x^1 \frac{dy}{y}\,
P\!\left(\frac{x}{y}\right) f(y,\mu^{2}).
\end{equation}
Here, $\alpha_s(\mu^{2})$ is the QCD strong coupling constant and $y$ is the momentum fraction of the parent parton before the splitting. $P_{qq}$, $P_{qg}$, $P_{gq}$ and $P_{gg}$ are the fundamental splitting functions of the DGLAP evolutions. Furthermore, from the initial gluon and sea-quark distributions, one can obtain their behavior at larger scales using the initial quark PDF. 
\par Further, our evolved valence quark, gluon, and sea-quark PDFs at $5$ GeV$^2$ have been compared with available theoretical extraction of global collaboration results of JAM21 \cite{Barry:2021osv}, xFitter \cite{Novikov:2020snp}, GRV \cite{Gluck:1991ey}, and MAP23 \cite{Pasquini:2023aaf} in Fig. \ref{fig:pion_comparison}. For the valence quark PDF, our NLO and NNLO results exhibit similar behavior at both low and high $x$, indicating the quality of our results. However, at $x \sim 1$, our results for the valence quark PDFs are smoothly decreasing distributions rather than linear, slightly faster distributions as in the above predictions. A higher distribution is observed at high $x$ region in the case of LO PDF evolutions, indicating that they carry a higher longitudinal momentum fraction compared to NLO and NNLO evolutions. The LO-evolved PDF is also found to have a lower distribution than other results in the low $x$ region. In Fig. \ref{fig:pion_comparison} (b, c), we have also compared our results of gluon $(xg(x,\mu^2))$ and sea-quark $(xS(x,\mu^2))$ distributions at all orders with the theoretical extraction results. The gluon distribution is found to have a larger magnitude than the extracted results, whereas the sea-quark distribution is observed to lie within a similar range as the extracted distributions. The singlet sea-quark distributions for the pion case, with inclusion of all possible quark-antiquark, are calculated as,
\begin{equation}
\begin{aligned}
S(x,\mu^{2}) =\;
&\sum_{q = u, d, s, c, b}
\left[ q(x,\mu^{2}) + \bar{q}(x,\mu^{2}) \right] \\
&- \left[ u_{val}(x,\mu^{2}) + \bar{d}_{val}(x,\mu^{2}) \right].
\end{aligned}
\end{equation}
However, the production of top quark-antiquark pairs inside the pion is highly suppressed due to the very large top-quark mass. Therefore, the top flavor contribution is neglected in the present work. Here, $u_{val}(x,\mu^{2})$ and $\bar{d}_{val}(x,\mu^{2})$ are the valence quark PDFs of the pion. Also, for the sea-quark and antiquark follows the symmetry,
\begin{equation}
\begin{aligned}
u_{\mathrm{sea}}(x,\mu^2) &= \bar{u}_{sea}(x,\mu^2), \quad
\bar{d}_{\mathrm{sea}}(x,\mu^2) = d_{sea}(x,\mu^2), \\
s(x,\mu^2) &= \bar{s}(x,\mu^2), \quad
c(x,\mu^2) = \bar{c}(x,\mu^2), \\
b(x,\mu^2) &= \bar{b}(x,\mu^2).
\end{aligned}
\end{equation}
Both the gluon and sea-quark distributions are found to dominate at the low $x$-region, while valence quarks dominate at the high $x$. The NNLO evolved PDF is found to have a lower distribution compared to the LO and NLO in the case of gluons, but vice versa is observed in the case of sea-quark distributions. Both the gluons and sea-quark distributions are found to vanish in the region $x \ge 0.6$. Another observation to be made is that the sea-quark distribution of all orders is found to have a smoothly decreasing function, which was not seen in the theoretical extractions. 
\par We have also calculated the average momentum fraction $\langle x \rangle$ carried by the valence quark-antiquark, gluons, and sea-quarks at higher scales. These contributions have been plotted with respect to energy scales $\mu^2$ in the region $\mu^2=1$ to $10^3$ GeV$^2$ in Fig. \ref{fig:kaon_results} (a). The  $\langle x \rangle$ of valence quark-antiquark is found to decrease with an increase in energy scales, indicating the splitting of valence quarks into gluons. While the average momentum fraction of gluon and sea-quarks is found to increase with an increase in energy scales. We have also observed that the total gluon and sea-quarks carry higher momentum fraction than the valence quarks in the scales $\mu^2 \ge 5$ GeV$^2$. At each scale, the valence, gluon, and sea quarks obey the sum rule of 
\begin{equation}
\int_{0}^{1} dx\, x 
\left[
f_{val}(x,\mu^{2})
+\bar{f}_{val}(x,\mu^{2})
+ S(x,\mu^{2})
+ g(x,\mu^{2})
\right]
=1.
\end{equation}
\begin{table*}
\caption{Comparison of the lowest Mellin moments $\langle x^n \rangle$ up to $n=4$ of the pion valence PDF at different scales.}
\label{moments_tab_new}
\centering
\renewcommand{\arraystretch}{1.15}
\begin{tabular}{c c c c c c}
\hline\hline
 & $\mu^2$ (GeV$^2$) & $\langle x \rangle$ & $\langle x^2 \rangle$ & $\langle x^3 \rangle$ & $\langle x^4 \rangle$ \\
\hline

\multicolumn{6}{c}{\textbf{Low scale ($\mu^2 \sim 1.69$ GeV$^2$)}} \\
\hline
DSE-RL~\cite{Bednar:2018mtf} & 1.69 & 0.268 & 0.125 & 0.076 & 0.054 \\
WI-An ~\cite{Bednar:2018mtf} &  & 0.268 & 0.114 & 0.059 & 0.037 \\
JAM fit ~\cite{Barry:2018ort} &  & 0.268 & 0.127 & 0.074 & 0.048 \\
JAM DY ~\cite{Barry:2018ort} &  & $0.30 $ & -- & -- & -- \\
MAP~\cite{Pasquini:2023aaf} &  & $ 0.29 \pm 0.015 $ & -- & -- & -- \\
xFitter ~\cite{Novikov:2020snp} &  & $ 0.275 \pm 0.03$ & -- & -- & -- \\
BLFQ-NJL \cite{Lan:2019rba}  &  & $0.271^{+0.020}_{-0.020}$ & $0.124^{+0.014}_{-0.014}$ & $0.069^{+0.009}_{-0.009}$ & $0.044^{+0.007}_{-0.007}$ \\
This Work  &  & $0.283^{+0.025}_{-0.012}$ & $0.123^{+0.017}_{-0.008}$ &$0.065^{+0.011}_{-0.005}$  & $0.038^{+0.008}_{-0.004}$  \\
\hline
\multicolumn{6}{c}{\textbf{Intermediate scale ($\mu^2 \sim 4, \, 5.76$ GeV$^2$)}} \\
\hline
Lattice-3 \cite{Bar:2016jof} & 4 &  & -- & -- & -- \\
	Sutton ~\cite{Sutton:1991ay}&    &$0.24\pm0.01$&$0.10\pm0.01$&$0.058\pm0.004$& \\
	Hecht ~\cite{Hecht:2000xa}& &$0.24$&0.098&0.049& \\
	Chen ~\cite{Chen:2016sno}&    &$0.26$&$0.11$&$0.052$& \\
    xFitter ~\cite{Novikov:2020snp} &  & $  0.25 \pm 0.025$ & -- & -- & -- \\
    Han \cite{Han:2018wsw} &  & $  0.255\pm 0.015$ & -- & -- & -- \\
    MAP~\cite{Pasquini:2023aaf} &  & $ 0.26 \pm 0.015 $ & -- & -- & -- \\
    GRVPI1 \cite{Gluck:1991ey} &  & $0.195$ & -- & -- & -- \\
    Ding \cite{Ding:2019lwe} &  & $ 0.24 \pm 0.015$ & -- & -- & -- \\
	BSE ~\cite{Shi:2018mcb}& &0.24& & &\\
	QCDSF/UKQCD  [lattice QCD]~\cite{QCDSF-UKQCD:2006nvr}&  &$0.27\pm0.01$&$0.13\pm0.01$&$0.074\pm0.010$&\\
	DESY  [lattice QCD]~\cite{Abdel-Rehim:2015owa} & &$0.214\pm0.015$ & &  &\\
	ETM  [lattice QCD]~\cite{Oehm:2018jvm} & &$0.207\pm0.011$&$0.163\pm0.033$& &\\
	JAM fit~\cite{Barry:2018ort}& &$0.245\pm 0.005$&$0.108\pm0.003$&&\\
	BLFQ-NJL \cite{Lan:2019rba}& &$0.245^{+0.018}_{-0.018}$&$0.106^{+0.012}_{-0.012}$&$0.057^{+0.008}_{-0.008}$&$0.035^{+0.005}_{-0.005}$\\
This Work  &  &$0.246^{+0.025}_{-0.012}$ &$0.099^{+0.025}_{-0.012}$  &$0.049^{+0.007}_{-0.003}$  &$0.028^{+0.005}_{-0.002}$  \\
Detmold ~\cite{Detmold:2003tm} & 5.76 & $0.24\pm0.01$ & $0.09\pm0.03$ & $0.043\pm0.015$ & -- \\
BLFQ-NJL \cite{Lan:2019rba}&  &$0.236^{+0.018}_{-0.018}$&$0.101^{+0.011}_{-0.011}$&$0.054^{+0.007}_{-0.007}$&$0.032^{+0.005}_{-0.005}$\\
 xFitter ~\cite{Novikov:2020snp} &  & $  0.24 \pm 0.025$ & -- & -- & -- \\
This Work  &  &$0.237^{+0.017}_{-0.006}$ & $0.094^{+0.011}_{-0.004}$ &$0.046^{+0.006}_{-0.002}$  &$0.026^{+0.004}_{-0.001}$  \\
\hline
\multicolumn{6}{c}{\textbf{Higher scales ($\mu^2 \sim 27, \, 49$ GeV$^2$)}} \\
\hline

Watanabe ~\cite{Watanabe:2017pvl} & 27 & 0.23 & 0.094 & 0.048 & -- \\
Nam ~\cite{Nam:2012vm} &  & $0.214^{+0.016}_{-0.030}$ & $0.087^{+0.010}_{-0.019}$ & $0.044^{+0.006}_{-0.011}$ & $0.026^{+0.004}_{-0.008}$ \\
MAP~\cite{Pasquini:2023aaf} &  & $ 0.225 \pm 0.015 $ & -- & -- & -- \\
Wijesooriya ~\cite{Wijesooriya:2005ir} &  & $0.217\pm0.011$ & $0.087\pm0.005$ & $0.045\pm0.003$ & -- \\
xFitter ~\cite{Novikov:2020snp} &  & $ 0.21 \pm 0.02$ & -- & -- & -- \\
This Work  &  &$0.205^{+0.013}_{-0.003}$ &$0.075^{+0.007}_{-0.002}$  &$0.035^{+0.004}_{-0.001}$  &$0.019^{+0.003}_{-0.001}$  \\
Sutton ~\cite{Sutton:1991ay} & 49 & $0.200\pm0.015$ & $0.080\pm0.007$ & -- & -- \\
BLFQ-NJL\cite{Lan:2019rba}&  & $0.202^{+0.015}_{-0.015}$ & $0.079^{+0.009}_{-0.009}$ & $0.040^{+0.005}_{-0.005}$ & $0.023^{+0.003}_{-0.003}$ \\
xFitter ~\cite{Novikov:2020snp} &  & $  0.205 \pm  0.02$ & -- & -- & -- \\
This Work  &  &$0.195^{+0.012}_{-0.002}$ &$0.070^{+0.007}_{-0.001}$  &$0.032^{+0.040}_{-0.001}$  &$0.017^{+0.020}_{-0.00}$  \\
\hline\hline
\end{tabular}
\end{table*}
\begin{table*}[t]
\centering
\caption{First Mellin moments ($\langle x \rangle$) of gluon and sea-quark distributions at different energy scales $\mu^2$, compared with global analyses of xFitter \cite{Novikov:2020snp}, JAM \cite{Barry:2018ort} and MAP \cite{Pasquini:2023aaf}.}
\renewcommand{\arraystretch}{1.2}
\setlength{\tabcolsep}{6pt} 
\resizebox{\textwidth}{!}{%
\begin{tabular}{c c c c c c c c c}
\hline\hline
$\mu^2$ (GeV$^2$) 
& \multicolumn{2}{c}{This Work} 
& \multicolumn{2}{c}{xFitter \cite{Novikov:2020snp}} 
& \multicolumn{2}{c}{JAM \cite{Barry:2018ort}} 
& \multicolumn{2}{c}{MAP \cite{Pasquini:2023aaf}} \\
\cline{2-3} \cline{4-5} \cline{6-7} \cline{8-9}
& $\langle x \rangle_g$ & $\langle x \rangle_{\text{sea}}$
& $\langle x \rangle_g$ & $\langle x \rangle_{\text{sea}}$
& $\langle x \rangle_g$ & $\langle x \rangle_{\text{sea}}$
& $\langle x \rangle_g$ & $\langle x \rangle_{\text{sea}}$ \\
\hline
1.69  & $0.36^{+0.016}_{-0.043}$ & $0.08^{+0.006}_{-0.017}$  & $0.26 \pm 0.15$ &$0.19 \pm 0.16$&$0.30\pm 0.02$ & $0.16\pm 0.02$ & $0.33\pm 0.06$ & $0.09\pm 0.04$ \\
4.00  & $0.40^{+0.010}_{-0.032}$ & $0.10^{+0.004}_{-0.016}$ & $0.25\pm 0.13$ & $0.25\pm 0.13$ & -- & -- & $0.37\pm 0.05$ & $0.11 \pm 0.03$ \\
5.00  & $0.41^{+0.009}_{-0.030}$ & $0.11^{+0.003}_{-0.016}$ & $0.26 \pm 0.13$ & $0.25 \pm 0.12$ & $0.35\pm 0.02$ & $0.17\pm 0.01$ & $0.37\pm 0.05$ & $0.12\pm 0.03$ \\
10.00 & $0.43^{+0.007}_{-0.026}$ & $0.13^{+0.003}_{-0.016}$ & $0.31\pm 0.06$ & $0.22 \pm 0.08$ & $0.37 \pm 0.02$ & $0.19 \pm 0.01$ & $0.39 \pm 0.05$ & $0.13 \pm 0.02$ \\
27.00 & $0.44^{+0.005}_{-0.022}$ & $0.14^{+0.002}_{-0.016}$ & $0.32 \pm 0.10$ & $0.25 \pm 0.10$ & -- & -- & $0.40 \pm 0.04$ & $0.15 \pm 0.02$ \\
\hline\hline
\end{tabular}
}
\label{mellingluon}
\end{table*}
\begin{figure*}[t] 
    \centering
    \includegraphics[width=\textwidth]{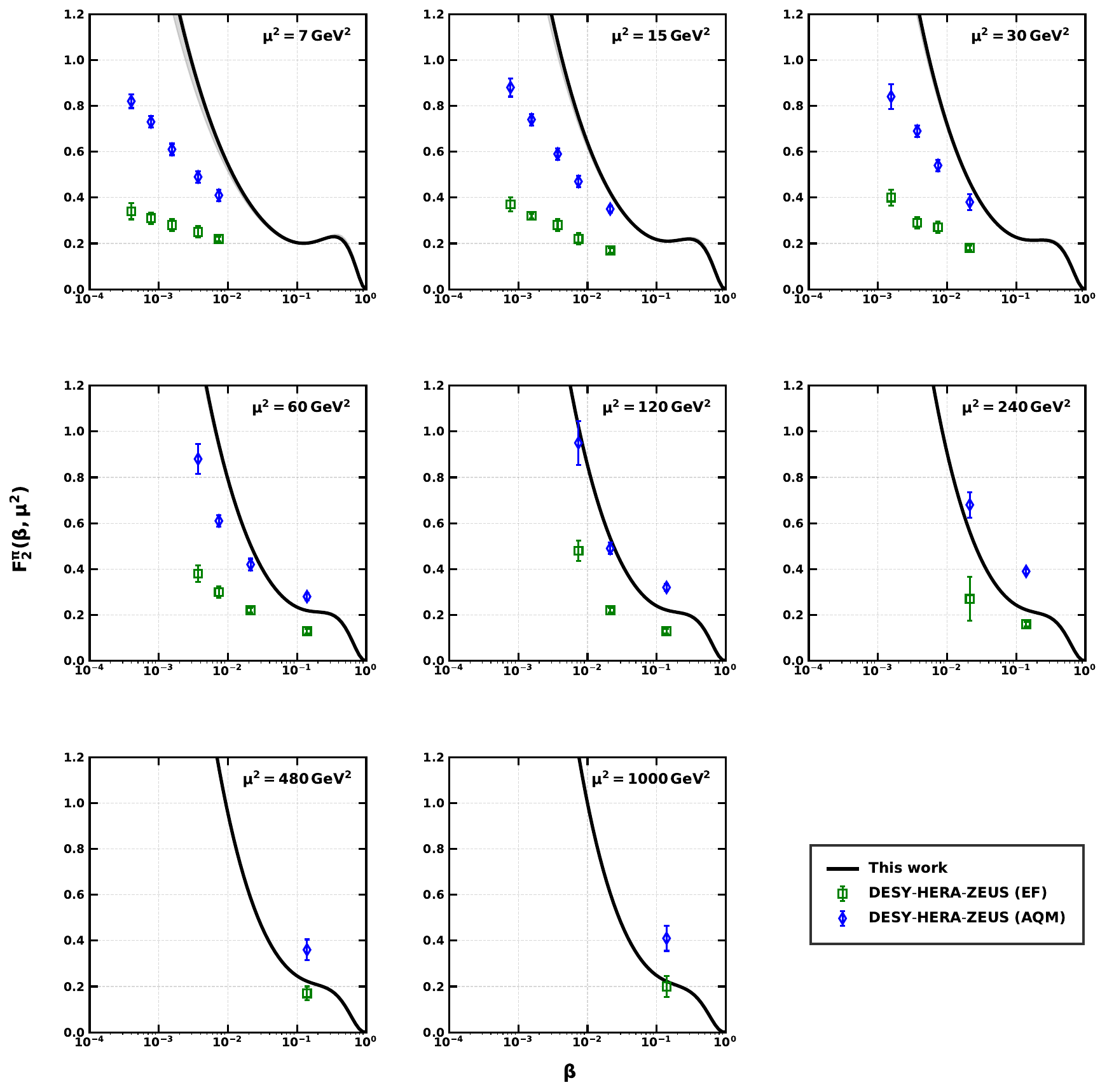}
    \caption{
    (Color online) Structure function $F_2^\pi(\beta, \mu^2)$ for the pion as a function of $\beta$ at fixed experimental values of $\mu^2$. The data points are taken from the ZEUS collaboration at DESY-HERA \cite{ZEUS:2002gig}. The solid black line indicates the central value and the gray shaded bands indicate the uncertainty resulting from the variation of the initial scale $\mu_{0} = 0.6 \pm 0.1\text{ GeV}$.
    }
    \label{fig:f2_structure_function_full_width}
\end{figure*}
In Fig. \ref{fig:kaon_results} (b), we have plotted the Mellin moment $\langle x^n \rangle$ of the valence quark antiquark distribution as a function of $\mu^2$ up to $n=4$. Here, we compare the average momentum fraction carried by the valence quarks with available theoretical extraction results of JAM21 \cite{Barry:2021osv}, xFitter \cite{Novikov:2020snp}, GRV \cite{Gluck:1991ey}, and MAP23 \cite{Pasquini:2023aaf}.Our results match those of the global extractions except for the GRV results. Additionally, the numerical values of the lowest four moments of the valence quark PDFs have been compared with the available phenomenological model \cite{Bednar:2018mtf,Lan:2019rba,Chen:2016sno,Shi:2018mcb,Ding:2019lwe,Detmold:2003tm,Watanabe:2017pvl,Nam:2012vm,Wijesooriya:2005ir}, lattice simulations results \cite{Bar:2016jof,Sutton:1991ay,Hecht:2000xa,Han:2018wsw,Ding:2019lwe,QCDSF-UKQCD:2006nvr,Abdel-Rehim:2015owa,Oehm:2018jvm} and theoretical extraction results    \cite{Barry:2018ort,Barry:2021osv,Pasquini:2023aaf,Gluck:1991ey,Gluck:1999xe,Novikov:2020snp} in Table. \ref{moments_tab_new}. The Mellin moments are found to match all other results. We observed that at $\mu^2=49$ GeV$^2$, the valence quark-antiquark is found to carry only $39\%$ of the total momentum fraction of the pion, the rest $45\%$ and $16\%$ carried by the gluon and sea-quarks, respectively, by taking the central initial scale $\mu_0^2=0.36$ GeV$^2$. The average momentum fractions carried by gluons and sea quarks at different scales are presented in the Table. \ref{mellingluon}. While comparing with the theoretical global extraction results of MAP \cite{Pasquini:2023aaf}, xFitter \cite{Novikov:2020snp}, MAP \cite{Barry:2018ort}, our average momentum fraction carried by the gluon is found to be higher, while the momentum fraction carried by the sea-quarks is found to be consistent with MAP collaboration results and deviates from the other results. This indicates that the higher Fock-state contributions are needed to study the pion internal structure. Overall, our predictions are found to be in good agreement with the theoretical extraction results.
\begin{figure*}[t] 
    \centering
    \includegraphics[width=\textwidth]{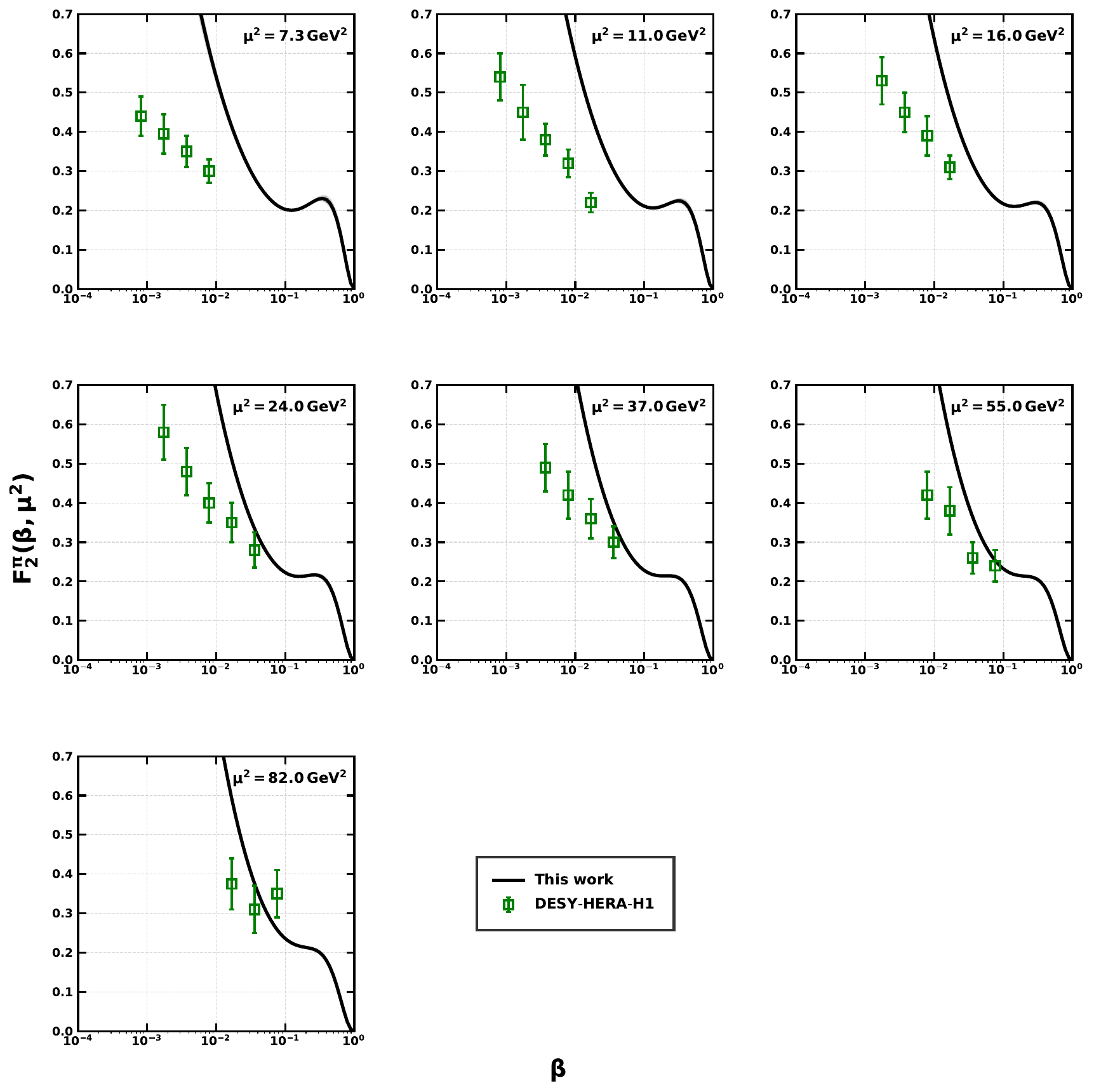}
    \caption{
    (Color online) Structure function $F_2^\pi(\beta, \mu^2)$ for the pion as a function of $\beta$ at various experimental scales $\mu^2$. The data are taken from the H1 collaboration in DESY-HERA \cite{H1:2010hym}. The solid black line indicates the central value while the gray shaded bands indicate the uncertainty resulting from the variation of the initial scale $\mu_{0} = 0.6 \pm 0.1\text{ GeV}$.
    }
    \label{fig:f2_h1_comparison}
\end{figure*}

\begin{figure*}[!t] 
    \centering
    \begin{subfigure}[b]{0.48\textwidth}
        \centering
        \includegraphics[width=\textwidth]{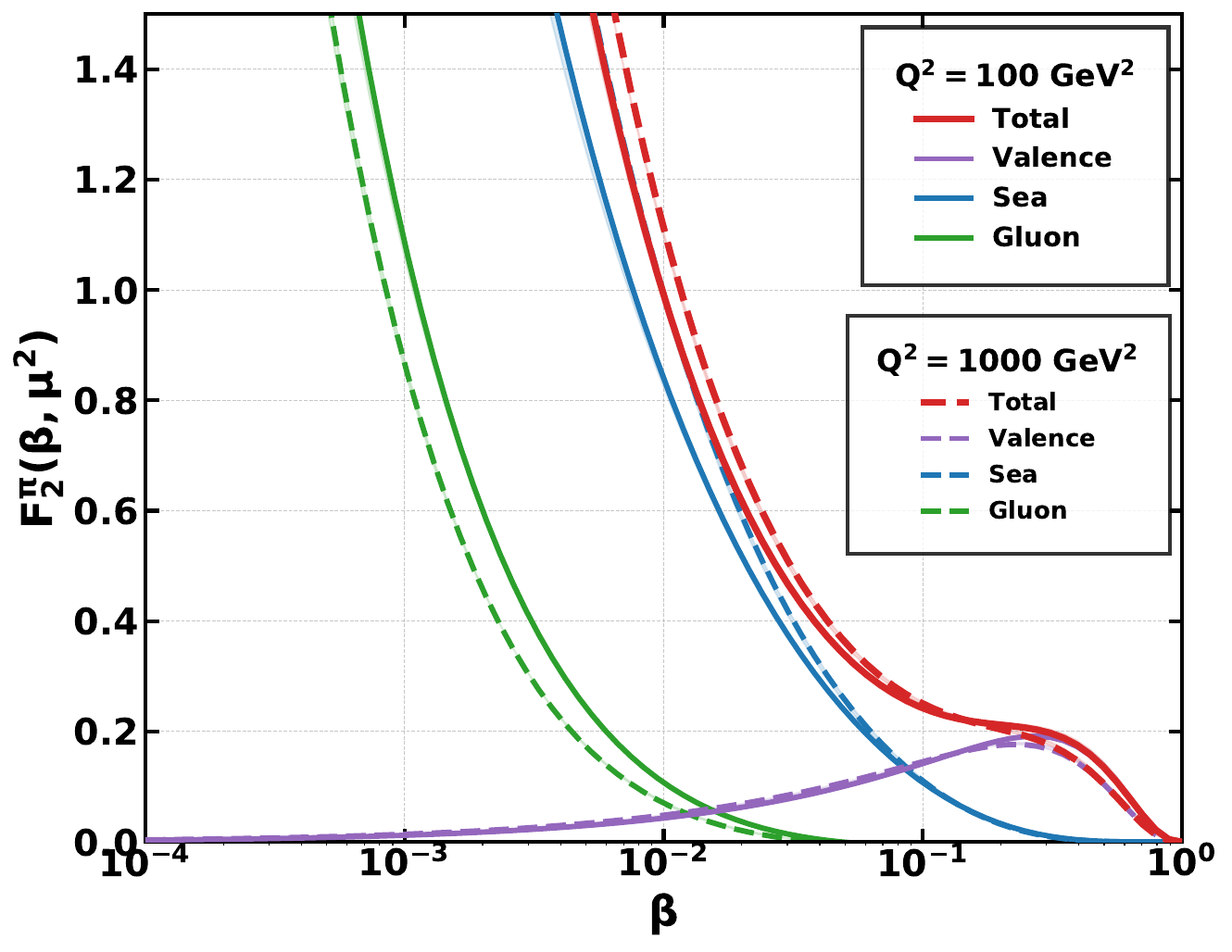}
        \caption{}
        \label{fig:F2contribution1}
    \end{subfigure}
    \hfill 
    \begin{subfigure}[b]{0.48\textwidth}
        \centering
        \includegraphics[width=\textwidth]{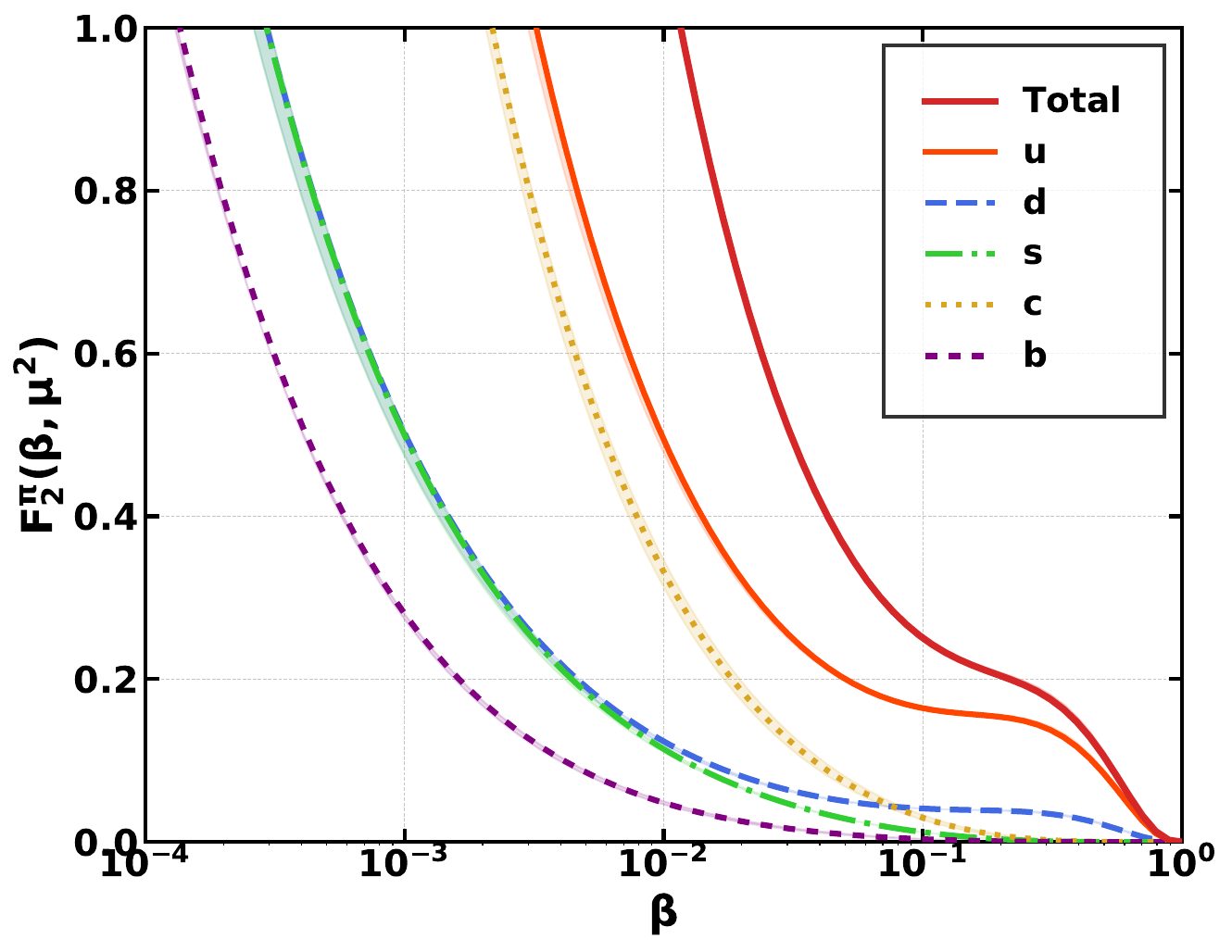}
        \caption{}
        \label{fig:F2contribution2}
    \end{subfigure}
    \caption{(Color online)(a) Contributions to the $F_2^{\pi}$ from the valence quark, sea quark and gluon distribution at energy scales $\mu^2 = 100 \text{ GeV}^2$ and $\mu^2 = 1000 \text{ GeV}^2$  (b) contributions to the $F_2^{\pi}$ from different quark flavors at $\mu^2 = 1000 \text{ GeV}^2$. The error bands represent the results of this work, uncertainty from the initial scale $\mu_{0} = 0.6 \pm 0.1\text{ GeV}$}
    \label{fig:F2_contributions}
\end{figure*}

\begin{figure}[t]
    \centering
    \includegraphics[width=1\columnwidth]{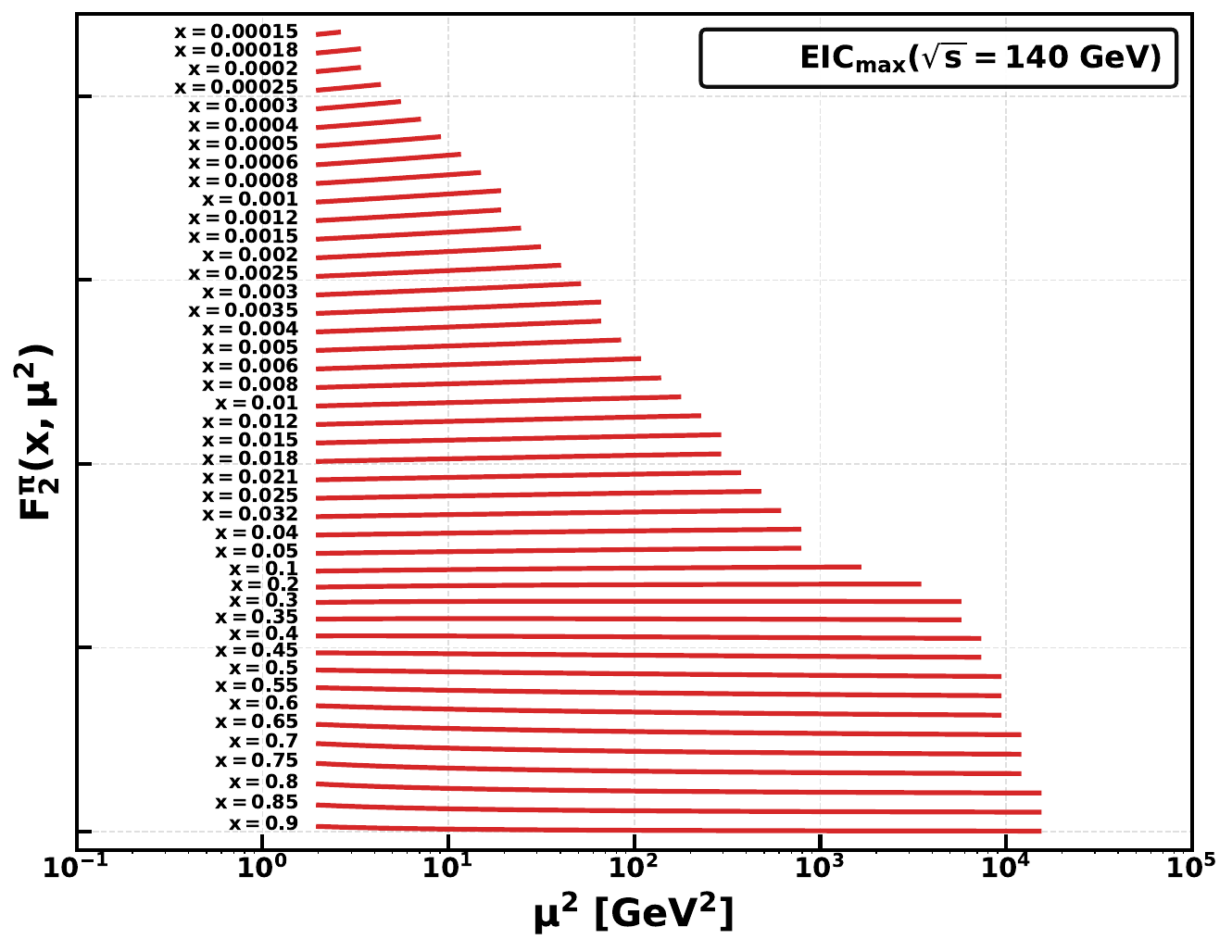}
    \caption{
    (Color online) The pion structure function $F_2^{\pi}$ as a function of $\mu^2$ for different values of $x$. The kinematic limit is applied according to the upcoming EIC with a maximum center-of-mass energy (${\sqrt{s_{max}} = \text{140 GeV}}$), by taking the initial scale uncertainty of the order $\mu_{0} = 0.6 \pm 0.1\text{ GeV}$ of our initial PDF.
    }
    \label{fig:pion_f2_decomposed}
\end{figure}
\begin{figure*}[!t] 
    \centering
    \begin{subfigure}[b]{0.48\textwidth}
        \centering
        \includegraphics[width=\textwidth]{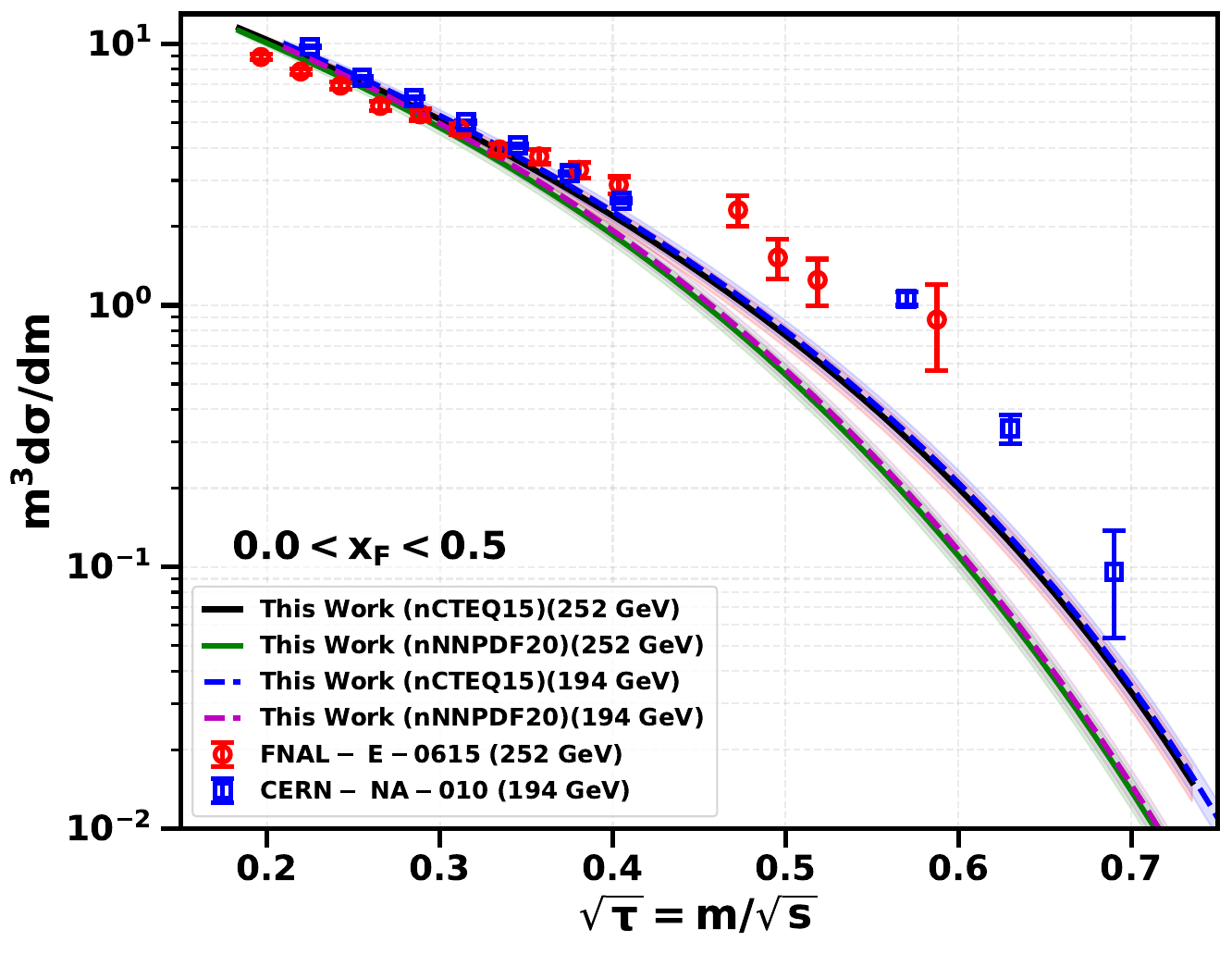}
        \caption{}
        \label{fig:e615na10}
    \end{subfigure}
    \hfill 
    \begin{subfigure}[b]{0.48\textwidth}
        \centering
        \includegraphics[width=\textwidth]{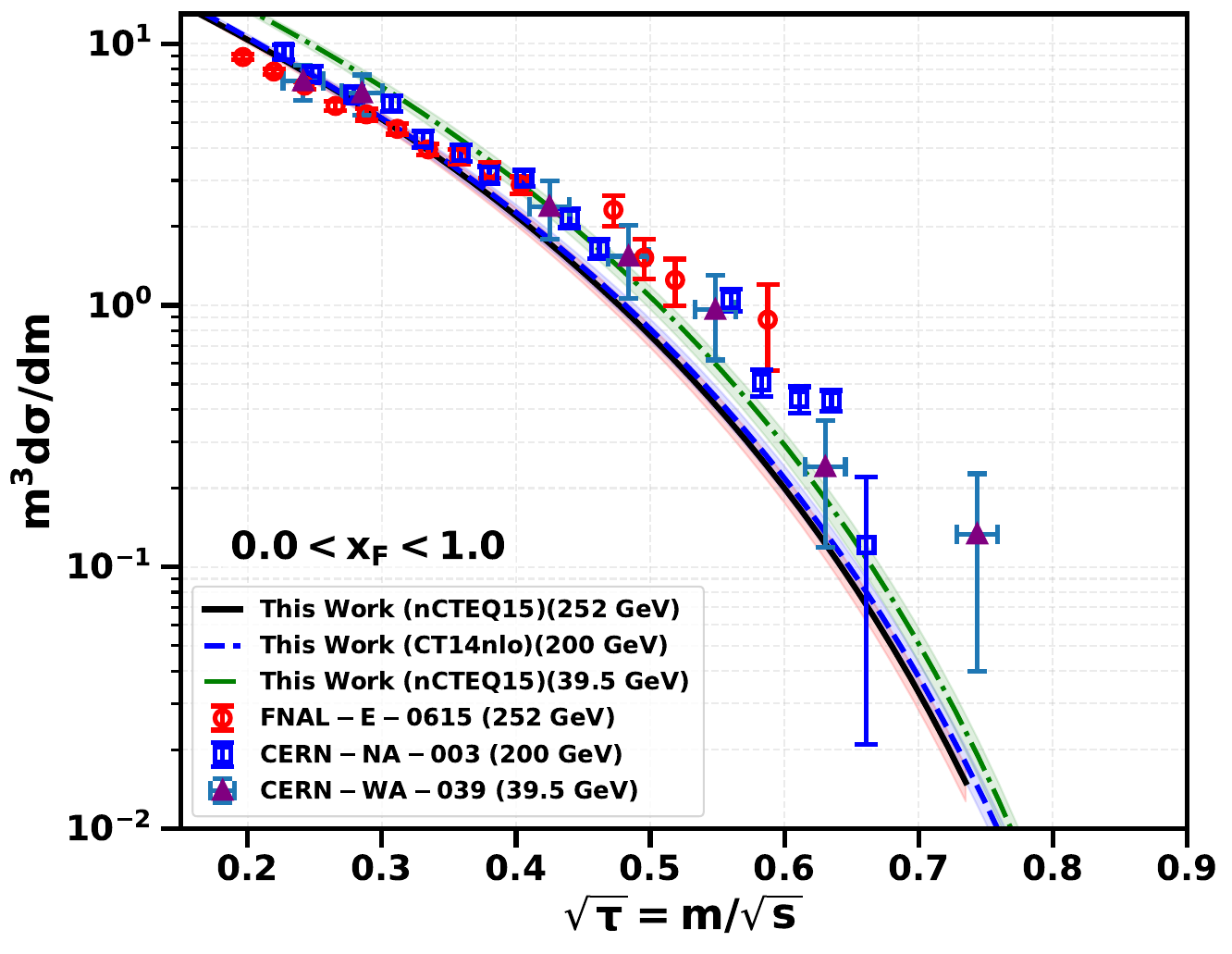}
        \caption{}
        \label{fig:e615na003}
    \end{subfigure}
    \caption{(Color online)(a) The cross-section $m^3d\sigma/dm$ for the Drell-Yan process as a function of $\sqrt{\tau}$ have been compared with available experimental data. The data of the FNAL-E-0615 \cite{E615:1989bda}, CERN-NA-003 \cite{NA3:1983ejh}, and CERN-NA-010 \cite{NA10:1985ibr} experiments with center of mass energy 252 GeV, 200 GeV, and 194 GeV, respectively. The FNAL-E-0615 and CERN-NA-010 data correspond to tungsten data, while the CERN-NA-003 data correspond to a platinum target. The error bands represent the results of this work, including the uncertainty from the initial scale $\mu_{0} = 0.6 \pm 0.1\text{ GeV}$.  }
    \label{fig:full_moments_figure}
\end{figure*}
\begin{figure*}[!t] 
    \centering
    \begin{subfigure}[b]{0.50\textwidth}
        \centering
        \includegraphics[width=\textwidth]{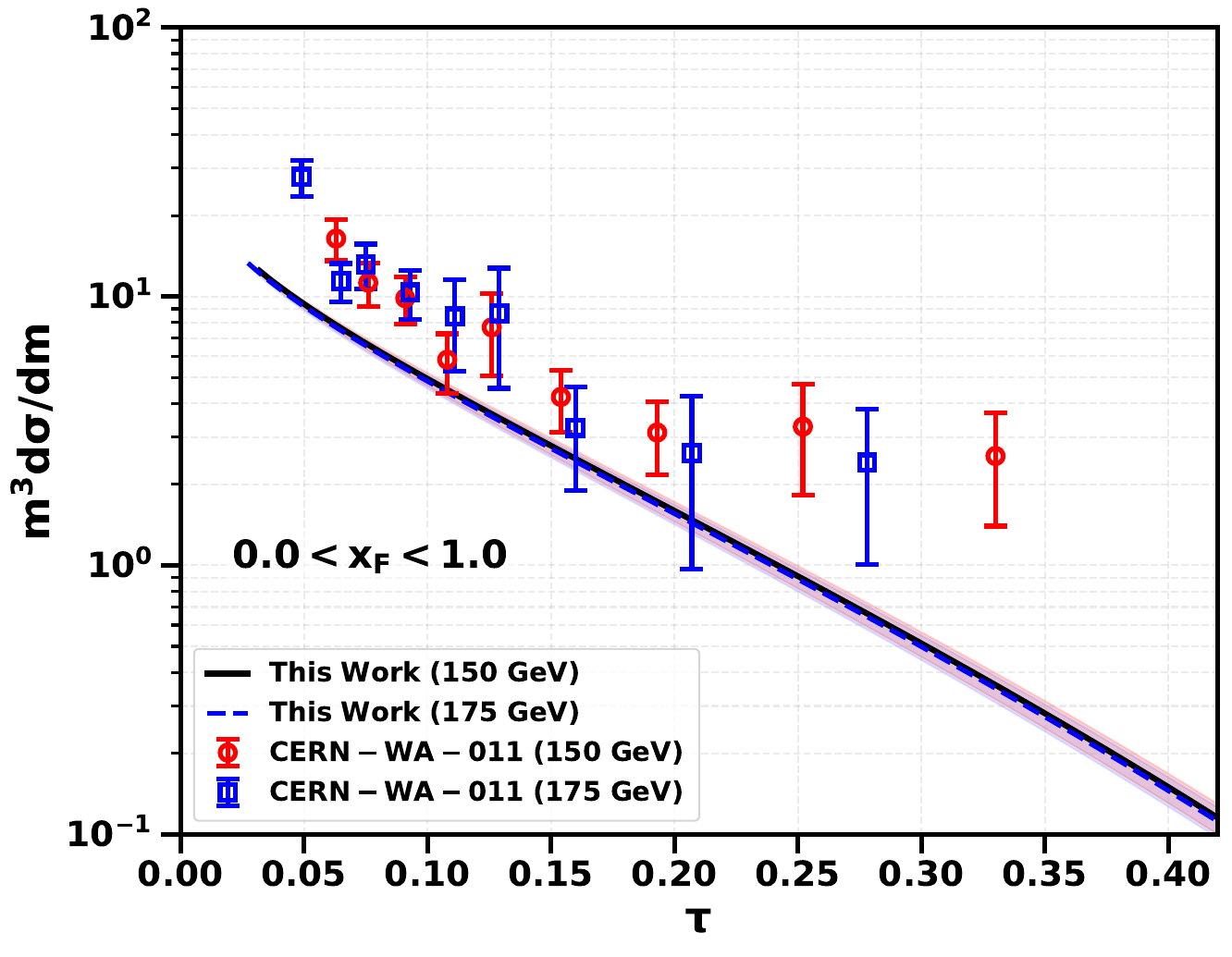}
        \caption{}
        \label{fig:e615na10}
    \end{subfigure}
    \hfill 
    \begin{subfigure}[b]{0.49\textwidth}
        \centering
        \includegraphics[width=\textwidth]{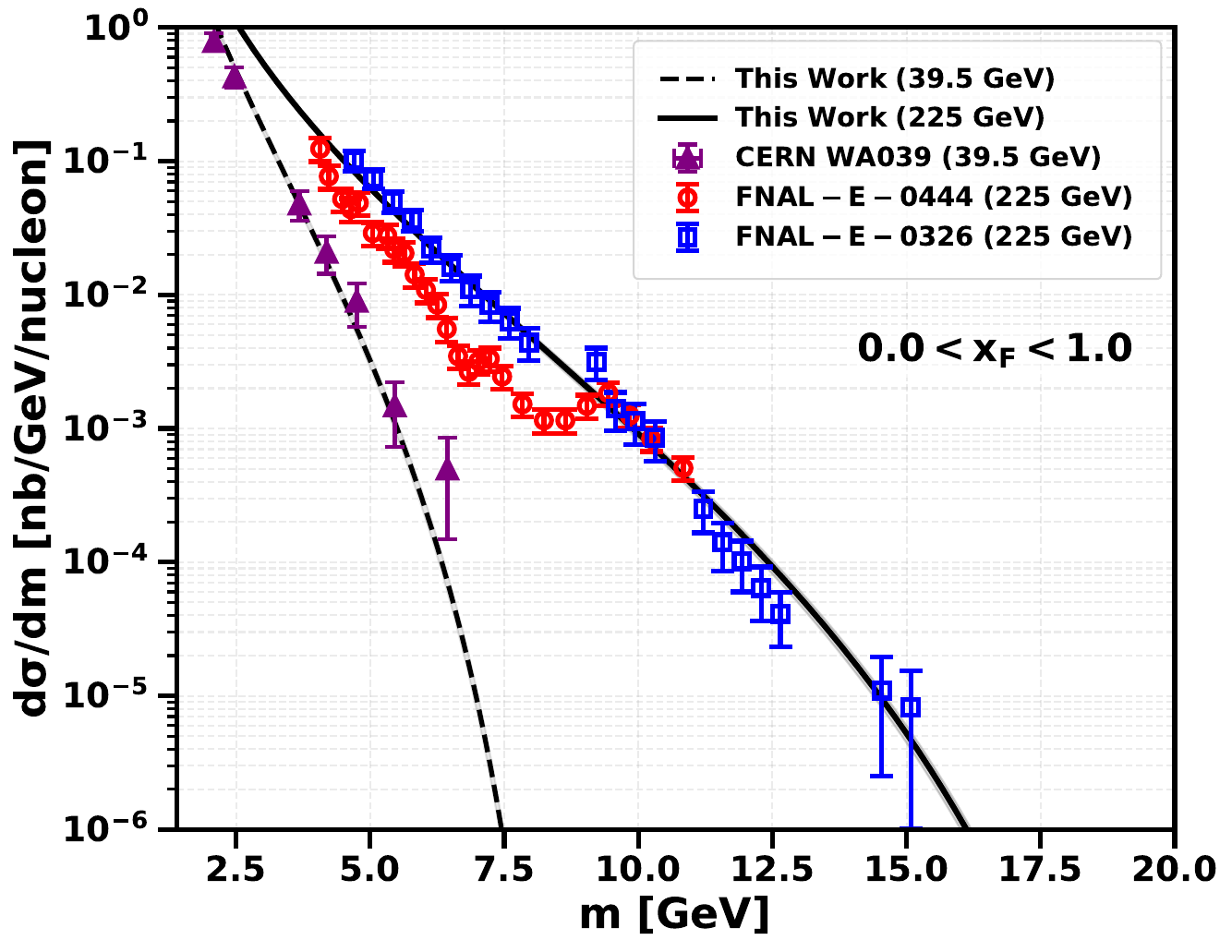}
        \caption{}
        \label{fig:e615na003}
    \end{subfigure}
    \caption{(Color online) (a) The cross-section $m^3d\sigma/dm$ for the Drell-Yan process as a function of $\tau$ have been compared with available experimental data. The data of the CERN-WA-011 experiments with center of mass energy 150 GeV and 175 GeV, respectively, are taken from Ref. \cite{Barate:1979da}. (b) The cross-section $d\sigma/dm$ for the Drell-Yan process as a function of $m$. The data from CERN-WA-039, FNAL-E-0326, and FNAL-E-0444 experiments with 39.5 GeV and 225 GeV, respectively, are taken from \cite{Corden:1980xf,Greenlee:1985gd,Newman:1979tv}. The FNAL-E-0326 and the CERN-WA-039 data correspond to a tungsten target, while the FNAL-E-0444 data correspond to a carbon target. The error bands represent the results of this work, including the uncertainty from the initial scale $\mu_{0} = 0.6 \pm 0.1\text{ GeV}$ }
    \label{fig:crosssections}
\end{figure*}
\section{$F_2$ Structure Function}
\label{sfpdf}
The direct extraction of PDFs from experiments is challenging due to the lack of stable pion targets, which limits the experimental constraints. Nevertheless, relatively richer information is available for the pion structure function $F_{2}^{\pi}$ over a wide kinematic range. In particular, the pion $F_{2}^{\pi}$ structure function has been investigated through leading-neutron electro-production measurements $e + p \rightarrow e' + n + X$, performed at HERA in 2002 and 2009 \cite{ZEUS:2002gig,H1:2010hym}. So from the PDF evolutions, we have calculated the NLO structure function in perturbative QCD using the same initial scale. The NLO pion structure function can be calculated as \cite{Lan:2019rba}
\begin{equation}
\begin{aligned}
F_{2}^{\pi}(\beta, \mu^{2}) 
= \sum_{q} e_{q}^{2}\, \beta \Bigg\{ 
& f(\beta, \mu^{2}) + \bar f(\beta, \mu^{2})  + \frac{\alpha_{s}(\mu^{2})}{2\pi} 
 \\
& \times \Big[ 
C_{q} \otimes \big( f(\beta, \mu^{2}) + \bar f(\beta, \mu^{2}) \big)  \\
& \qquad\qquad + 2\, C_{g} \otimes g(\beta, \mu^{2}) 
\Big] 
\Bigg\},
\end{aligned}
\end{equation}
with
\begin{align}
C_{q}[z] &= \frac{4}{3} \left[ \frac{1+z^{2}}{1-z} \left( \ln \frac{1-z}{z} - \frac{3}{4} \right) \right]_{+}, \\
C_{g}[z] &= \frac{1}{2} \left[ (z^{2} + (1-z)^{2}) \ln \frac{1-z}{z} - 1 + 8z(1-z) \right].
\end{align}
Here, $q$ and $e_q$ represent the flavor index and electric charge of the quark flavor $q$ (in units of elementary charge), respectively. $\beta=x_\pi$ is the Bjorken variable of the pion. In leading-neutron experiments $\beta$ is calculated as $\beta=\frac{x_p}{1-x_L}$, where $x_p$ is the parton momentum
fraction relative to the proton and $x_L$ is the momentum fraction
carried by the neutron relative to the proton. In this work, we have taken $x_L=0.73$ as done in HERA \cite{ZEUS:2002gig,H1:2010hym}. $z$ is the hard-scattering momentum fraction of the parton. We have compared our results of $F_2^{\pi}(\beta,\mu^2)$ structure functions with available DESY--HERA--ZUES \cite{ZEUS:2002gig} and DESY--HERA--H1 \cite{H1:2010hym} experimental data in Figs. \ref{fig:f2_structure_function_full_width} and \ref{fig:f2_h1_comparison}. The two ZUES datasets correspond to different pion fluxes used to determine the $F_2$ structure functions. These are the additive quark model (AQM) and effective one-pion-exchange flux (EF) \cite{Lan:2019rba}. 
\par In Fig. \ref{fig:f2_structure_function_full_width}, we have compared our pion structure function $F_{2}^{\pi}$ with both AQM and EF results at different experimental energy scales. We observe that our results are found to have slightly higher distribution compared to both the AQM and EF results up to $\mu^2=7$ to $30$ GeV$^2$. Our results are found to consistent with the AQM data for the scales $\mu^{2} = 60$, $120$ and $240$ GeV$^2$. Beyond this scale, our predictions show good agreement with the EF data. At $\mu^{2} =1000~\mathrm{GeV}^{2}$, our results exactly coincide with the central value of the single available EF data point, while they deviate from the AQM datasets. This indicates the need for more experimental measurements to better constrain the pion structure functions. At low energy scales, we observe a peak around $x=0.5$ that decreases as the energy scale increases. This kind of behavior was also observed in the BLFQ-NJL model \cite{Lan:2019rba}. We have also compared our results with DESY--HERA--H1 data \cite{H1:2010hym} in Fig. \ref{fig:f2_h1_comparison}. At $\mu^2=7.3$, $11$ and $16$ GeV$^2$, our results disagree with all the data points, showing a higher distribution compared to all. However, for other energies, our results matched several data points across different datasets, indicating the overall reliability of our LCQM predictions. We have also observed that the sea-quark contributions are coming higher compared to valence and gluon for the $F_2^{\pi}(\beta,\mu^2)$ as shown in Fig. \ref{fig:F2_contributions} (a). The bottom, charm, and strange quark contributions are found to increase with increasing energy scales, indicating more gluon splitting into sea-quarks, which results in the increase in distribution at low $\beta$ region. The individual contributions have been plotted in Fig. \ref{fig:F2_contributions} (b) for all the quark flavours of the pion after the evolutions at $\mu^2=1000$ GeV$^2$. We have also plotted the evolved $F_2^{\pi}(x,\mu^2)$ with respect to the energy scales at different values of $x$ in Fig. \ref{fig:pion_f2_decomposed}. The  $F_2^{\pi}(x,\mu^2)$ structure function is found to be less sensitive to $\mu^2$ in the high $x$ region, while more sensitive at low $x$. The kinematic cuts have been applied with the maximum EIC energy (${\sqrt{s_{max}} = \text{140 GeV}}$) . Our LCQM results agree with all the other results, more experimental input is required for the pion structure functions. The ongoing COMPASS/Amber at CERN \cite{Moinester:2000eq,Grube:2015fac} and electroproduction experiments in JLab \cite{CLAS:2005vxa} are measuring the pion structure functions. Upcoming EIC will provide the structure function data of pion through the Sullivan process over a wide range of energies \cite{Aguilar:2019teb}.
\begin{figure*}[t] 
    \centering
    \includegraphics[width=\textwidth]{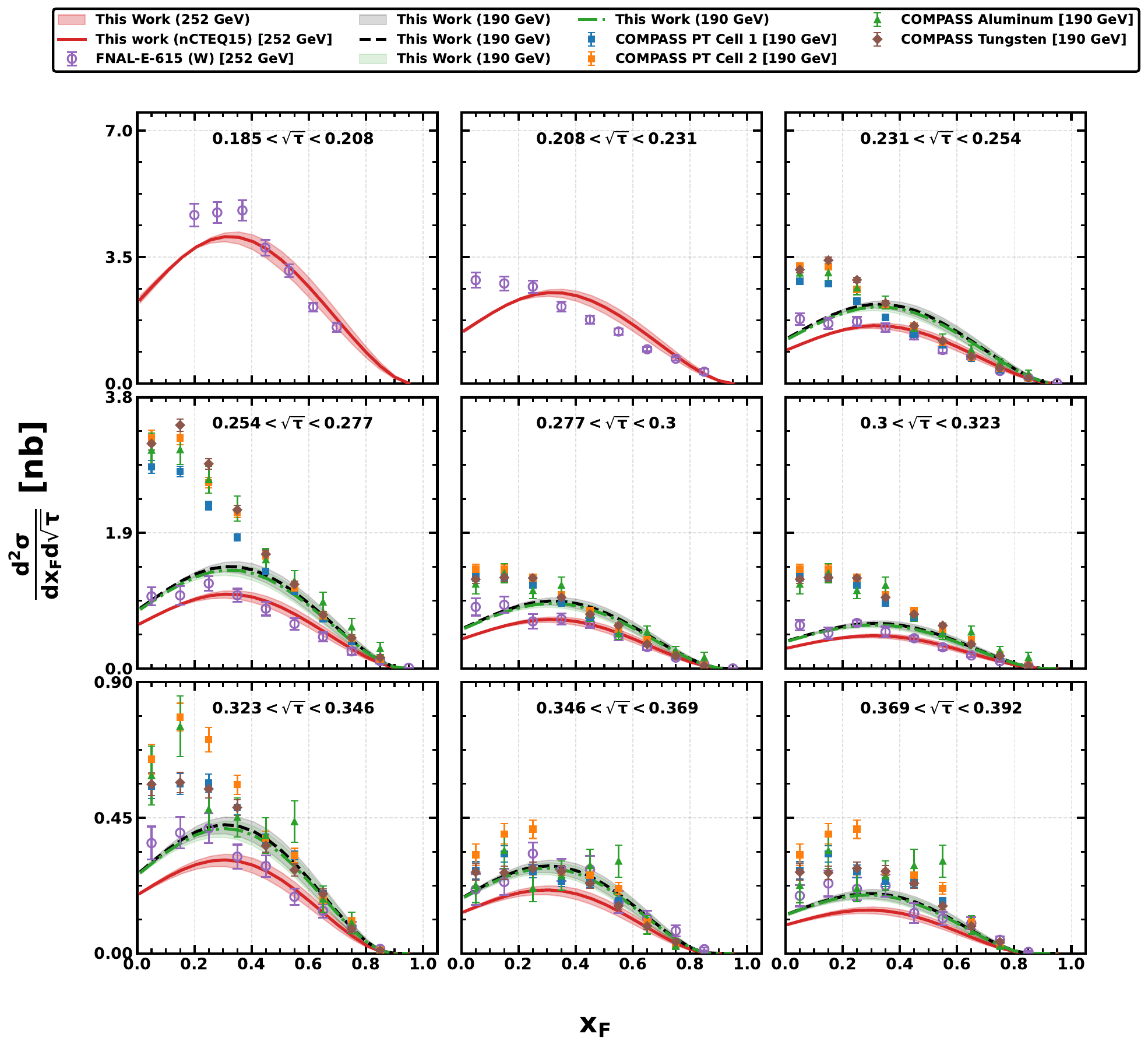}
    \caption{
    (Color online) The pion induced Drell-Yan cross-section $d^2\sigma/dx_F d\sqrt{\tau}$ of this work as a function of $x_F$ compared with FNAL-E-0615 \cite{E615:1989bda} and COMPASS-II \cite{Meyer-Conde:2019frd}. The error bands represent the results of this work, including the uncertainty from the initial scale $\mu_{0} = 0.6 \pm 0.1\text{ GeV}$.
    }
    \label{compasscomparison}
\end{figure*}

\begin{figure*}[t] 
    \centering
    \includegraphics[width=\textwidth]{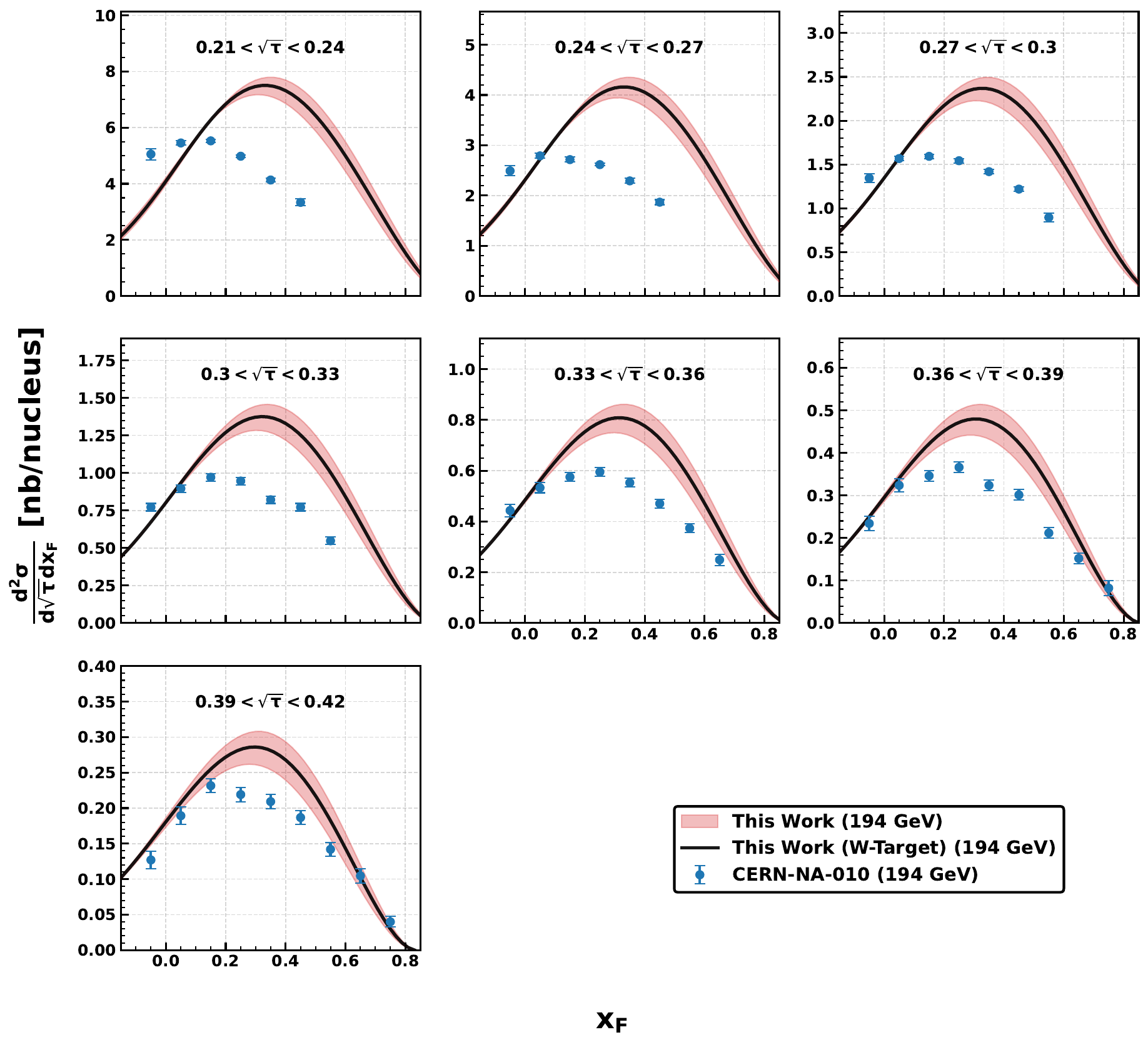}
    \caption{
    (Color online) The pion-induced Drell-Yan cross-section $d^2\sigma/dx_F d\sqrt{\tau}$ of this work as a function of $x_F$ compared with CERN-NA-010 \cite{NA10:1985ibr}. The error bands represent the results of this work, including the uncertainty from the initial scale $\mu_{0} = 0.6 \pm 0.1\text{ GeV}$.
    }
    \label{na10comparison}
\end{figure*}

\begin{figure*}[t] 
    \centering
    \includegraphics[width=\textwidth]{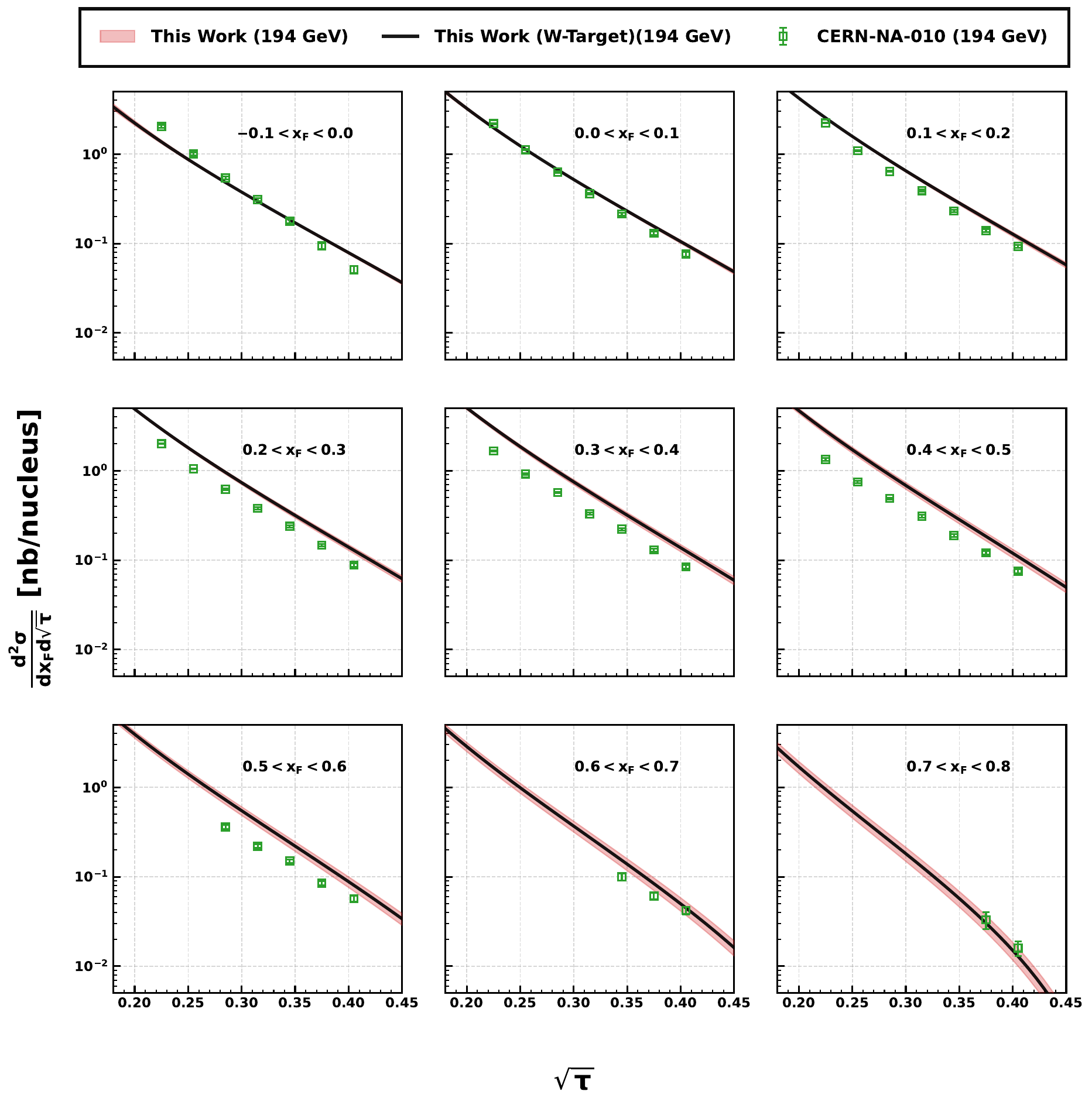}
    \caption{
    (Color online) The pion-induced Drell-Yan cross-section $d^2\sigma/dx_F d\sqrt{\tau}$ of this work as a function of $\sqrt{\tau}$ compared with CERN-NA-010 \cite{NA10:1985ibr}. The error bands represent the results of this work, including the uncertainty from the initial scale $\mu_{0} = 0.6 \pm 0.1\text{ GeV}$.
    }
    \label{na10comparisonagain}
\end{figure*}

\begin{figure*}[t] 
    \centering
    \includegraphics[height=1\textheight, width=0.85\textwidth, keepaspectratio]{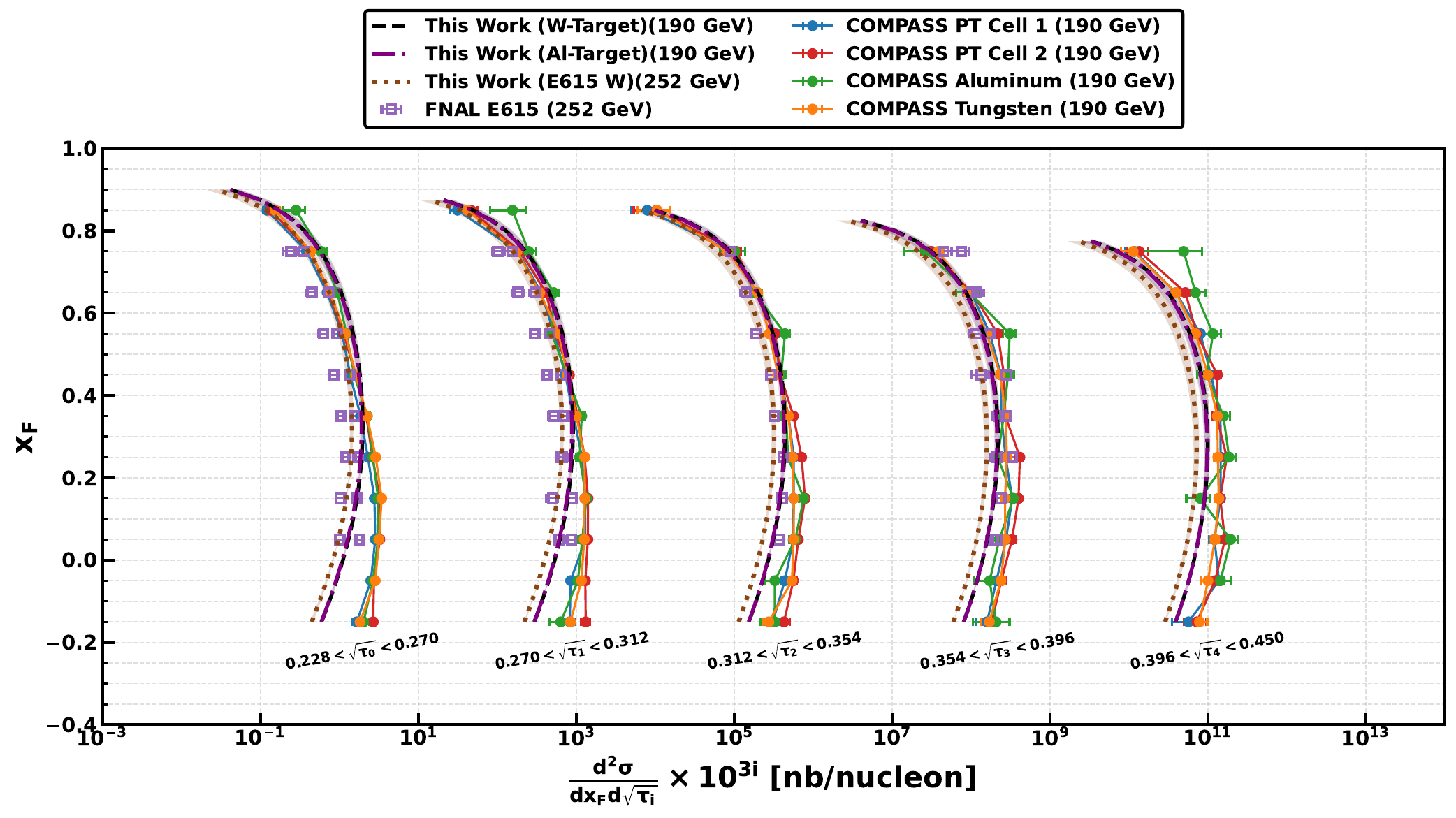}
    \caption{(Color online) The pion induced Drell-Yan cross-section $d^2\sigma/dx_F d\sqrt{\tau}$ of this work as a function of $x_F$ in the bin of $\sqrt{\tau_i}$ with experimental cross-section data from COMPASS-II (190-GeV) \cite{Meyer-Conde:2019frd} and FNAL-E-0615 (252-GeV) \cite{E615:1989bda}. }
    \label{fig:compass12_large}
\end{figure*}

\begin{figure*}[t] 
    \centering
    \includegraphics[height=1\textheight, width=0.8\textwidth, keepaspectratio]{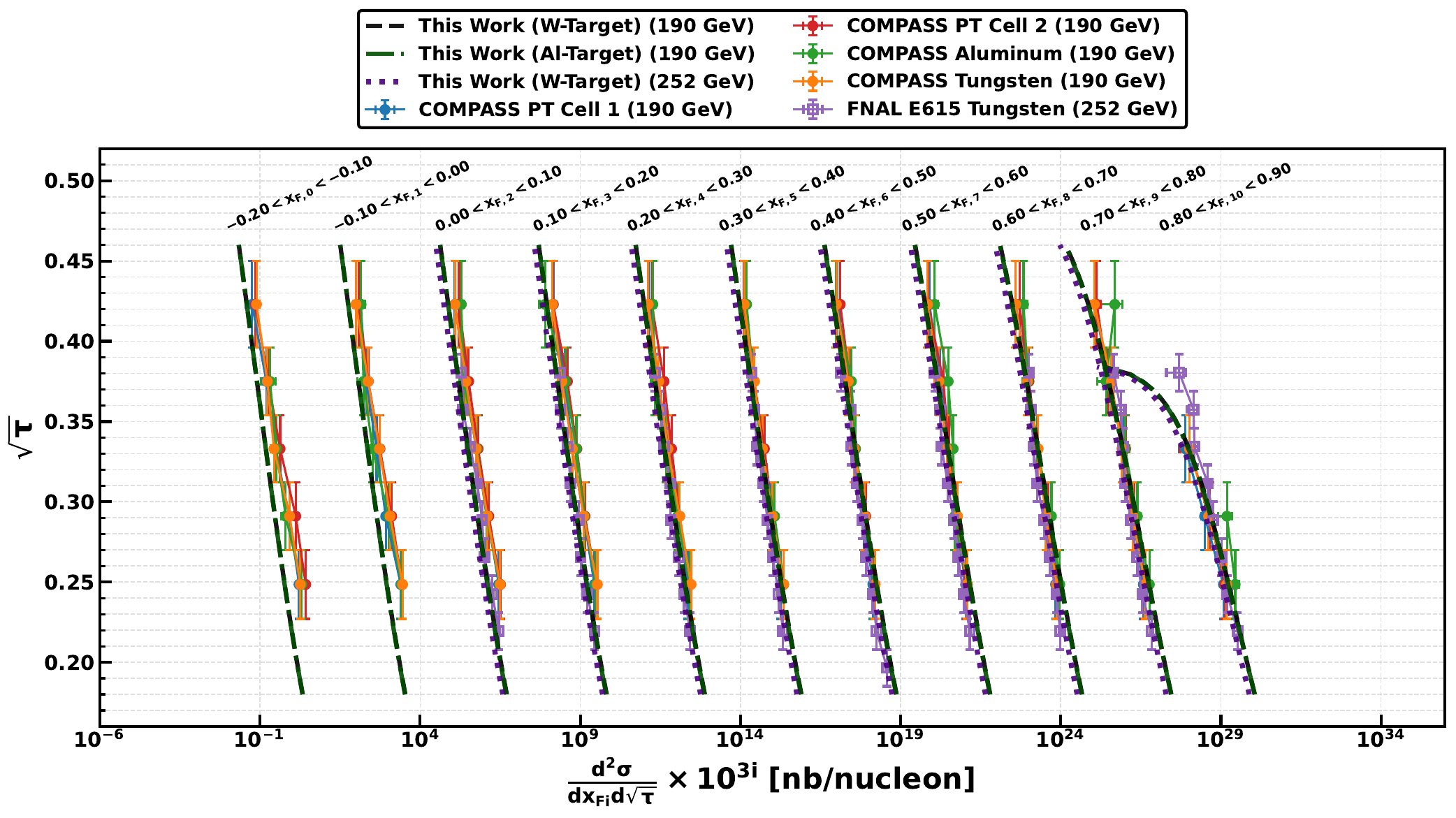}
    \caption{(Color online) The pion induced Drell-Yan cross-section $d^2\sigma/dx_F d\sqrt{\tau}$ of this work as a function of $\sqrt{\tau}$ in the bins of $x_{F_i}$ with experimental cross-section data from COMPASS-II (190-GeV) \cite{Meyer-Conde:2019frd} and FNAL-E-0615 (252-GeV) \cite{E615:1989bda}}
    \label{fig:compass13_large}
\end{figure*}

\section{unpolarized Drell-Yan Cross-Section}
\label{crosssection}
In this section, we have determined the theoretical pion-induced Drell-Yan cross-section using our initial valence quark PDF in the LCQM, and perform an extensive comparison against several datasets, including the recent results from COMPASS-II experiments \cite{Meyer-Conde:2019frd}. The pion-induced Drell-Yan process have been studied in several fixed target dilepton production experiments $\pi^{\pm} + A \rightarrow l^{+}l^{-} + X$ (A=target nucleus, all the experiments are done for di-muon productions), including FNAL-E-0444 \cite{Newman:1979tv}, CERN-WA-011 \cite{Barate:1979da}, CERN-WA-039 \cite{Corden:1980xf},  FNAL–E-0326 \cite{Greenlee:1985gd}, CERN-NA-010 \cite{NA10:1985ibr}, CERN-NA-003 \cite{NA3:1983ejh} and FNAL-E-0615 \cite{E615:1989bda}. Most of the experiments have been performed by colliding $\pi^-$ with a tungsten target at different energies. We define $l^+$ and $l^-$ as the momenta of the two outgoing leptons. The Drell-Yan process is described by the lepton-pair invariant mass $Q$, which represents the mass of the produced lepton pair, the center-of-mass square energy $s$, which represents the square of the total energy available, the Feynman variable $x_F$, the hadronic scaling variable $\tau$, the rapidity $Y$, and the partonic scaling variable $z$ and $y$. These kinematics are related as 
\begin{align}
    s &= (p_1 + p_2)^2, & q &= l^+ + l^-, \\
    Q^2 &= q^2, & Y &= \frac{1}{2} \ln \frac{q_0 + q_3}{q_0 - q_3}, \displaybreak \\
    x_{\text{F}} &= x_a - x_b, & \tau &= \frac{m^2}{s}, \\
    z &= \frac{m^2}{\hat{s}} = \frac{\tau}{x_1 x_2}, & y &= \frac{\frac{x_a}{x_b}e^{-2Y} - z}{(1-z)(1 + \frac{x_1}{x_2}e^{-2Y})},
\end{align}
where $x_a$ and $x_b$ denote the Bjorken-$x$ or the fraction of the hadron momentum $p_1$ and $p_2$ carried by the annihilating parton (or antiparton), and can be written in terms of rapidity $Y$ and scaling variable $z$,
\begin{align}
    x_a &= \sqrt{\frac{\tau}{z} \frac{1 - (1-y)(1-z)}{1 - y(1-z)}} e^{Y}, \\
    x_b &= \sqrt{\frac{\tau}{z} \frac{1 - y(1-z)}{1 - (1-y)(1-z)}} e^{-Y}. 
\end{align}
The cross-section in terms of the target nucleus and incoming pion PDFs at NLO can be given by \cite{Meyer-Conde:2019frd, Lan:2019rba,Anastasiou:2003yy},
\begin{equation}
\begin{aligned}
    \frac{m^3 d^2\sigma}{dm \, dY} &= \frac{8\pi\alpha^2_s(\mu^{2})}{9} \frac{m^2}{s} \sum_{ij} \int dx_a dx_b \\
    &\times {w}_{ij}(x_1, x_2, s, m, \mu^2) f_{i/\pi}(x_1, \mu^2) f_{j/A}(x_2, \mu^2), 
\end{aligned}
\end{equation} 
where $w_{ij}$ is the hard-scattering kernels, which are expanded using the powers of the strong coupling constant $\alpha_s$, which we have taken the form as described in Ref. \cite{Anastasiou:2003yy}. The sum includes the $q\bar{q}$ annihilation channels, as well as quark-gluon ($qg$) and antiquark-gluon $(\bar{qg})$. Here $f_{i/\pi}(x_1, \mu^2)$ is the final evolved PDF and $f_{j/A}(x_2, \mu^2)$ is the nuclear PDFs. Furthermore, the pion-induced Drell-Yan cross-section is transformed in the form of Feynman scaling variable $x_F$ and scaling variable $\tau$,
\begin{equation}
\begin{aligned}
x_{a,b} &= \frac{x_F \pm \sqrt{x_F^{\,2} + 4\left(\frac{\tau}{z}\right)}}{2}, \\
Y &= \frac{1}{2}\ln \left(
\frac{\sqrt{x_F^{\,2} + 4\left(\frac{\tau}{z}\right)} + x_F}
     {\sqrt{x_F^{\,2} + 4\left(\frac{\tau}{z}\right)} - x_F}
\right).
\end{aligned}
\end{equation}
The cross section in terms of $x_F$ and $\tau$ is found to be 
\begin{equation}
\frac{d^{2}\sigma}{d\sqrt{\tau}\,dx_F}
=
\frac{\sqrt{s}}{\sqrt{x_F^{\,2}+4\frac{\tau}{z}}}\;
\frac{d^{2}\sigma}{dm\,dY},
\end{equation}
with the Jacobian transformation of 
\begin{equation}
\left|\frac{\partial(m,Y)}{\partial(\tau,x_F)}\right|
=
\frac{\sqrt{s}}{2\sqrt{\tau}\,\sqrt{x_F^{\,2}+4\frac{\tau}{z}}}.
\end{equation}
To evaluate the cross-section of pion-induced Drell-Yan experiments, we implement two different nuclear PDFs: nCTEQ 2015 \cite{Kovarik:2015cma} and nNNPDF20 \cite{AbdulKhalek:2020yuc}, at the experimental scale depending upon the di-lepton mass. While comparing the nuclear PDF sets, nCTEQ15 yields a superior description of the tungsten (W) data compared to nNNPDF20 for this work; accordingly, we use the nCTEQ15 framework for all subsequent theoretical predictions. This comparison of choice of nuclear PDFs has been presented in Fig. \ref{fig:full_moments_figure} (a), where the nCTEQ15 nuclear PDF is found to be closer to the experimental results than the nNNPDF20. After integrating out the $Y$ dependence of the differential cross-section $m^3d\sigma/dmdY$, we obtain our results plotted as a function of $\sqrt{\tau}$ in Fig. \ref{fig:full_moments_figure} (a) and (b). In Fig. \ref{fig:full_moments_figure} (a), we have compared our predicted cross-section of LCQM with FNAL-E-0615 at 252 GeV \cite{E615:1989bda} and CERN-NA-10 \cite{NA10:1985ibr} at 194 GeV in the limit $0< x_F <0.5$. In the region $0 \le \sqrt{\tau} \le 0.4$, our predictions were found to match exactly with the experimental data, whereas beyond that region they were lower than that of the experimental data. The $m^3d^2\sigma/dm$ predictions are also compared with FNAL-E-0615 at 252 GeV, CERN--NA--3 at 200 GeV \cite{NA3:1983ejh} and CERN--WA--039 at 39.5 GeV \cite{Corden:1980xf} in Fig. \ref{fig:full_moments_figure} (b), and found to be in good agreement with them in the whole $\sqrt{\tau}$ region. In Fig. \ref{fig:crosssections} (a), we have compared our $m^3d\sigma/dm$ results as a function of $\tau$ with the available CERN--WA--011 data at 150 and 175 GeV \cite{Barate:1979da}. We observe that our calculations show slightly lower distributions compared to the experimental data points. While in Fig. \ref{fig:crosssections} (b), we have compared our $d\sigma/dm$ as a function of dilepton invariant mass $m$ with available CERN--WA--039 at 39.5 GeV\cite{Corden:1980xf}, FNAL--E--0444 \cite{Newman:1979tv} and FNAL--E--0326 at 225 GeV \cite{Greenlee:1985gd}. Our results are in excellent agreement with these results. One of the most important thing to note is that for comparison with FNAL-E-0444 \cite{Newman:1979tv}, CERN--WA--011 \cite{Barate:1979da}, and  CERN--NA--3 \cite{NA3:1983ejh} datasets, we have used the carbon (C), Beryllium (Be), and platinum (Pt) nuclear PDFs from nCTEQ15, respectively. For others, we have used the tungsten (W) nuclear PDFs as done in experiments. 

\par In recent years, the ongoing COMPASS experiments at CERN have also provided the pion-induced Drell-Yan cross section at $190$ GeV \cite{Meyer-Conde:2019frd}. So, we have also compared our Drell-Yan cross section of $d^2\sigma/dx_Fd\sqrt{\tau}$ with both FNAL-E-0615 and COMPASS-II experiments as a function of $x_F$ in Fig. \ref{compasscomparison} in the fixed range of $\sqrt{\tau}$. The COMPASS-II pion-induced Drell-Yan data have been obtained at 190 GeV on a tungsten and aluminum target, as well as a polarized target labeled PT cell 1 and PT cell 2, both at 190 GeV. Most of the results are found to be in good agreement with all of them. We have also compared the same with the CERN--NA--10 experimental data \cite{NA10:1985ibr} in the range $0.21 < \sqrt{\tau} < 0.42$ in Fig. \ref{na10comparison}, while as a function of $\sqrt{\tau}$ in Fig. \ref{na10comparisonagain} in the range of $-0.1 < x_F < 0.8$. We observed that our LCQM results are found to match most of the data points in the CERN--NA--010 data \cite{NA10:1985ibr}. The pion-induced Drell-Yan cross-section $d^2\sigma/dx_Fd\sqrt{\tau}$ is plotted as a function of Feynman variable $x_F$ for distinct bins of scaling variable $\sqrt{\tau}$ in Fig. \ref{fig:compass12_large} and for distinct bins of Feynman variable $x_F$ in Fig. \ref{fig:compass13_large}. Both the results of our LCQM are found to match the experimental datasets at fixed values of dilepton mass. The simultaneous agreement of our pion model across both experimental datasets confirms the universality of the extracted pion PDFs.

\FloatBarrier
\section{Conclusion}
\label{conclusion}
In this work, we have calculated the valence quark PDFs of the pion by solving the quark-quark correlation functions in the LCQM. We have presented it at both model scale and higher energy scales through leading order (LO), next-to-leading order (NLO), and next-next-to-leading order (NNLO)  Dokshitzer–Gribov–Lipatov–Altarelli–Parisi (DGLAP) equations. Our results for the valence PDFs are found to be consistent with modified FNAL-E-0615 experimental data, along with other theoretical extraction results. From the initial valence PDFs, we have also predicted the gluon and sea-quark distributions, which were then compared with the theoretical extraction results of JAM, xFitter, MAP, and GRV. We have also calculated the lower and higher order Mellin moments of valence PDFs at the initial scale as well as at higher scales. Our calculated Mellin moments are found to be in good agreement with lattice simulations and other theoretical extraction results. We have also observed that only $38\%$ of the momentum fraction is carried by the valence quark at $49$ GeV$^2$, the rest is carried by gluon and sea-quarks.
\par From the initial valence PDFs, we have also predicted the $F_2$ structure functions at different energy scales. These structure functions have been compared with the leading-neutron electroproduction data of HERA. We have also predicted the evolved $F_2$ structure functions at different values of $x$. We have also calculated the NLO pion-induced Drell-Yan cross section as a function of various kinematic variables. For the nuclear PDFs, we have used the CTEQ collaboration data. We have also compared our theoretical model results with the recent COMPASS-II data. Overall, the valence PDFs of the LCQM are found to be in excellent agreement with the experimental data.
\par These results are most important for the ongoing COMPASS/Amber experiments on pion Drell-Yan experiments, along with future electron-ion colliders in the USA and China. For future work, we are targeting the addition of gluon contributions in the higher Fock states.

\section*{Acknowledgment}
S.P.\ would like to thank Prof.\ Fredrick Olness, Prof.\ Amanda M.\ Cooper-Sarkar, and Prof.\ Dave Soper for useful discussions during the \textit{2026 IITB-CFNS-CTEQ School on Perturbative QCD for the EIC}, held from 8--15 February 2026.
H.D.\ would like to thank the Science and Engineering Research Board (SERB), Anusandhan National Research Foundation, Government of India, for financial support under the SERB-POWER Fellowship (Ref.\ No.\ SPF/2023/000116).
%
%
%

\clearpage
\bibliographystyle{apsrev}  
\bibliography{ref} 
\end{document}